\documentclass{aastex6}
\usepackage{mkfig}

\begin{document}

\title{A {\em STEREO} Survey of Magnetic Cloud Coronal Mass Ejections Observed
  at Earth in 2008--2012}

\author{Brian E. Wood\altaffilmark{1}, Chin-Chun Wu\altaffilmark{1},
  Ronald P. Lepping\altaffilmark{2}, Teresa Nieves-Chinchilla\altaffilmark{2,3},
  Russell A. Howard\altaffilmark{1}, Mark G. Linton\altaffilmark{1},
  Dennis G. Socker\altaffilmark{1}}

\altaffiltext{1}{Naval Research Laboratory, Space Science Division,
  Washington, DC 20375, USA; brian.wood@nrl.navy.mil}
\altaffiltext{2}{Heliophysics Science Division, NASA Goddard Space
  Flight Center, Greenbelt, MD 20771, USA}
\altaffiltext{3}{Catholic University of America, Washington, DC 20064, USA}

\begin{abstract}

     We identify coronal mass ejections (CMEs) associated with magnetic
clouds (MCs) observed near Earth by the {\em Wind} spacecraft from
2008 to mid-2012, a time period when the two {\em STEREO} spacecraft
were well positioned to study Earth-directed CMEs.  We find 31 out of
48 {\em Wind} MCs during this period can be clearly
connected with a CME that is trackable in {\em STEREO} imagery all the
way from the Sun to near 1~AU.  For these events, we perform full 3-D
reconstructions of the CME structure and kinematics, assuming a flux
rope morphology for the CME shape, considering the full complement of
{\em STEREO} and {\em SOHO} imaging constraints.  We find that the
flux rope orientations and sizes inferred from imaging are not well
correlated with MC orientations and sizes inferred from the {\em Wind}
data.  However, velocities within the MC region are reproduced
reasonably well by the image-based reconstruction.  Our kinematic
measurements are used to provide simple prescriptions for predicting
CME arrival times at Earth, provided for a range of distances from
the Sun where CME velocity measurements might be made.  Finally,
we discuss the differences in the morphology and kinematics of CME
flux ropes associated with different surface phenomena (flares,
filament eruptions, or no surface activity).

\end{abstract}


\section{Introduction}

     The term ``coronal mass ejection'' (CME) was created to describe
visible eruptions of mass from the Sun observed in white-light
coronagraphic images of the outer solar corona
\citep{rt73,jtg74,ajh84,rah85,sk87}
Ever since their discovery, CMEs have been a very
active area of solar physics research, due in part to the recognition
that CMEs are the cause of most strong, nonrecurrent geomagnetic
storms \citep{jtg93}.  This research has been stimulated by a
succession of space-based coronagraphs, particularly
the Large Angle Spectrometric COronagraph (LASCO) instrument on
board the {\em SOlar and Heliospheric Observatory} (SOHO), which
has been in continuous operation near Earth's L1 Lagrangian point
since 1996 \citep{geb95}.

     However, remote observations are not the only way to study
the CME phenomenon.  Many spacecraft launched into interplanetary
space in the space age have carried plasma and field
instruments designed to study the solar wind.  Most notable among
those specially dedicated to studying the solar wind are {\em Wind}
\citep{rpl95,kwo95} and the {\em Advanced
Composition Explorer} (ACE) \citep{ecs98},
launched in 1994 and 1997, respectively, with both spacecraft
still currently monitoring the solar wind near L1.
The plasma and field instruments on such spacecraft can
study solar transients such as CMEs in~situ, with in~situ detections
of CMEs often referred to as ``interplanetary coronal mass ejections,''
or ICMEs.

     The CME/ICME distinction represents an implicit
acknowledgment that it is not always possible to clearly
connect a presumed ICME observed near Earth with a verified CME event
observed near the Sun \citep{gml99,ctr05,lj06}.
From in~situ data alone, there is no definitive
procedure for unambiguously distinguishing ICMEs from the background
solar wind and other transients such as corotating interaction
regions (CIRs) or stream interaction regions (SIRs).
Many ICMEs are characterized by a region of low plasma $\beta$ and
strong, rotating magnetic fields, and these properties define
an ICME subclass called ``magnetic clouds'' (MCs)
\citep{lb81,km86,lfb88,rpl90,vb98}.

     By far the most favored current interpretation of MCs is that
they are magnetic flux ropes (FRs), tube-shaped structures permeated
by a helical magnetic field, with legs that stretch back to the Sun
\citep{km86,cjf95,vb98}.  Evidence for magnetic connection with the
Sun comes from the frequent observations of counterstreaming electron
flows within such structures \citep{jtg87,igr96}, which is more easily
explained with a geometry with field lines bending back towards the
photosphere in both directions (as in the FR picture) than with the
field line stretching towards the distant reaches of the heliosphere
in one direction.  In principle, remote sensing should be better
suited for studying global CME morphology.  Coronagraphic imaging,
particularly from SOHO/LASCO, has been used to provide support for the
FR paradigm \citep{jc97,seg98,stw01,wbm04a,jk07}.  In particular,
large, bright CMEs seen in LASCO images are often found to have large
circular rims, which can be interpreted as outlining the apexes of FRs
viewed edge-on \citep{bew99,afrt06}.

     However, many CMEs present a more confusing appearance in images,
making morphological interpretation difficult.  By enabling the
simultaneous viewing of CMEs from multiple locations, the {\em Solar
TErrestrial RElations Observatory} (STEREO) mission was designed to
improve such analyses.  The two STEREO spacecraft, launched in
2006~October, were placed into orbits around the Sun similar to
Earth's, but with STEREO-A ahead of Earth in its orbit and STEREO-B
behind, and with both spacecraft gradually drifting away from Earth at
a rate of about $22^{\circ}$ per year.  Each STEREO spacecraft carries
four white light telescopes that observe at different distances from
the Sun \citep{raho08}.  There are two coronagraphs, COR1 and
COR2, that observe the Sun's white light corona at angular distances
from Sun-center of $0.37^{\circ}-1.07^{\circ}$ and
$0.7^{\circ}-4.2^{\circ}$, respectively, corresponding to distances in
the plane of the sky of $1.4-4.0$ R$_{\odot}$ for COR1 and
$2.5-15.6$~R$_{\odot}$ for COR2.  And there are two heliospheric
imagers, HI1 and HI2 \citep{cje09}, that monitor the
interplanetary medium in between the Sun and Earth, where HI1 observes
elongation angles from Sun-center of $3.9^{\circ}-24.1^{\circ}$ and
HI2 observes from $19^{\circ}-89^{\circ}$.

     Analyses of STEREO data have provided further evidence for a
flux rope morphology in CMEs, at least for some events
\citep{at09,cm09,bew09a}.  This has led
to the suggestion that FRs lie at the core of all CMEs \citep{av13}.
Only a fraction of ICMEs are perceived as MCs \citep{hvc03},
but this might be explained by noting that grazing
incidence events may present a messier in~situ signature that will not
be easily perceived as an MC, and for fast events the shock will
subtend a much larger solid angle than the presumed FR driver, meaning
that there should be many cases where an ICME shock is observed in~situ
without ever seeing the driver \citep{ng09}.

     The case for the FR paradigm could be greatly strengthened by
considering both in~situ and imaging constraints for a variety of
events and demonstrating conclusively that the FR shapes inferred
for these events from imaging are consistent with the
MC encounter times and FR orientations inferred in~situ.  The STEREO
spacecraft once again provide an excellent opportunity to do this,
especially given their heliospheric imaging capabilities, which
allow CMEs to be tracked all the way to 1~AU, where they can be
observed in~situ by ACE, {\em Wind}, or by the STEREO spacecraft
themselves.  A CME initiated on 2008~June~1 that ended up striking
STEREO-B on 2008~June~6 provided a perfect opportunity for such
an assessment, and for this event the inferences of FR size and
orientation from the imaging and in~situ data are in very good
agreement \citep{cm09,bew10a}.  However,
it is far from clear that this is the norm.  We here present
a survey of Earth-directed events that can address this issue
more fully.

     Our goal is to use STEREO and LASCO data to provide a full 3-D
kinematic and morphological analysis of a sample of CMEs observed as
MCs at Earth, specifically by {\em Wind}.  Aside from providing a
nice sample of events to study the characteristics of CMEs in the
STEREO era, we seek to address two questions in particular:  1. What
percentage of MCs detected at 1~AU can be definitively connected to
CMEs observed erupting from the Sun, and 2. When CME images are
analyzed in the context of the FR paradigm, are the inferred
FR structures consistent with the characteristics of the MCs
observed in~situ?

\section{Connecting Magnetic Clouds with Coronal Mass Ejections}

     For our study, we confine our attention to a time period
when the STEREO spacecraft were well located to observe Earth-directed
CMEs.  We want events well constrained by STEREO and SOHO/LASCO
imagery, and we want CMEs trackable out to
distances close to 1~AU in the HI2 field of view.  The requirement of
being able to track CMEs far from the Sun is in part simply due to a
desire to study CME kinematics far from the Sun, and in part because
tracking CMEs all the way to $\sim 1$~AU in images should maximize
our ability to connect an MC with its corresponding
CME.  It is somewhat arbitrary where to draw the line between a good
viewing geometry and a poor one, but in general the first year of
STEREO operations in 2007 was poor because the STEREO spacecraft were
still too close to Earth to stereoscopically study Earth-directed
events very well, and the post-2012 time period was generally poor
because the STEREO spacecraft had by then drifted too far ($>130^{\circ}$)
from the Sun-Earth line to effectively study and track Earth-directed
CMEs far from the Sun \citep{nl12b}.

     On 2012~July~12, STEREO-A reached a longitudinal distance of
$120^{\circ}$ from the Earth, which we decide to mark the end of our
survey time period, although STEREO-B would not reach this angular
distance from Earth for another 3 months.  As for the beginning of the
survey period, we decided to avoid 2007 but to consider all MC events
in 2008.  Thus, the first step in defining our sample of events was to
identify MCs observed by {\em Wind} between 2008~January~1 and
2012~July~12.  Our sources are the {\em Wind} MC lists from
\citet{rpl11,rpl15}, involving the use of both an
automated procedure \citep{rpl05} and visual inspection
to find MCs in the {\em Wind} database.  \citet{ccw15,ccw16}
have studied some statistical properties of these MC lists.

\begin{table}[t]
\scriptsize
\begin{center}
\caption{CMEs Associated with {\em Wind} MCs}
\begin{tabular}{ccccccccccccccc} \hline \hline
 & & & & & & & & & \multicolumn{4}{c}{Surface Activity} \\
\cline{10-13}
ID & CME Start & ICME Arrival &  $\phi_l$$^a$ & $\theta_l$$^a$ & Q$_l$$^a$ &
  $\phi_n$$^b$ & $\theta_n$$^b$ & Q$_n$$^b$ & Type$^c$ & Time & $\lambda_p$$^d$ & $\beta_p$$^d$ \\
   &           &              &      (deg)  &    (deg)       &           &
    (deg)  &   (deg)   &      &      &   & (deg)     &  (deg)    \\
\hline
1 & 2008-12-12T05:25 & 2008-12-16T08:00 & 65 & 1 & 2 & 68 & -10 & 2 & FE & 3:00 & 24 & 41 \\
2 & 2009-06-22T04:15 & 2009-06-27T12:00 & 52 & 42 & 3 & 257 & 65 & 3 & none &  &  &  \\
3 & 2009-07-15T04:30 & 2009-07-21T00:00 &297 & -17 & 2 & 335 & 4 & 3 & none &  &  &  \\
4 & 2009-09-03T01:00 & 2009-09-10T08:00 &247 & 57 & 2 & 226 & 27 & 2 & none &  &  &  \\
5 & 2009-09-25T12:00 & 2009-09-30T02:00 & 78 & 51 & 2 & 110 & 18 & 1 & none &  &  &  \\
6 & 2009-10-27T09:00 & 2009-11-01T04:00 &352 & 58 & 2 & 42 & 18 & 3 & none &  &  &  \\
7 & 2009-12-06T02:30 & 2009-12-12T06:00 &233 & -35 & 3 & 217 & -3 & 3 & none &  &  &  \\
8 & 2010-04-03T09:10 & 2010-04-05T09:00 &173 & 57 & 2 & 219 & -1 & 3 & FL(B7.4),FE & 9:04 & 2 & -21 \\
9 & 2010-05-23T16:35 & 2010-05-28T03:00 &125 & -81 & 1 & 78 & -70 & 2 & FL(B1.3),FE & 16:52 & 12 & 19\\
10 & 2010-06-16T06:35 & 2010-06-21T07:00 &260 & -4 & 2 & 266 & 32 & 3 & FE & 4:30 & 12 & -5 \\
11 & 2010-08-01T08:45 & 2010-08-03T18:00 &119 & -55 & 3 & 144 & -29 & 1 & FE & 6:30 & 17 & 29 \\
12 & 2010-09-11T01:10 & 2010-09-14T15:00 &282 & 64 & 3 & 52 & 74 & 3 & FE & 1:30 & -25 & 20 \\
13 & 2010-10-26T02:00 & 2010-10-30T10:00 &153 & 60 & 3 & 169 & 73 & 1 & FE & 0:30 & 1 & -55 \\
14 & 2010-12-14T05:00 & 2010-12-19T20:00 &319 & 24 & 3 & 285 & -19 & 3 & none &  &  &  \\
15 & 2011-02-15T01:55 & 2011-02-18T02:00 &11 & -4 & 3 & 53 & -72 & 3 & FL(X2.2),FE & 1:44 & 12 & -7 \\
16 & 2011-03-24T19:00 & 2011-03-29T16:00 &294 & 7 & 2 & 276 & 22 & 2 & none &  &  &  \\
17 & 2011-05-25T04:20 & 2011-05-28T07:00 &112 & -29 & 1 & 114 & -25 & 2 & FL(B1.9) & 3:41 & 12 & -16 \\
18 & 2011-06-02T07:45 & 2011-06-04T21:00 &317 & 20 & 3 & 274 & -17 & 2 & FL(C3.7),FE & 7:22 & -24 & -15\\
19 & 2011-06-14T06:10 & 2011-06-17T03:00 &18 & 66 & 3 & 208 & 17 & 3 & FE & 3:00 & -54 & -13 \\
20 & 2011-09-13T22:10 & 2011-09-17T04:00 &272 & -27 & 3 & 208 & 9 & 2 & FL & 22:00 & 3 & 10 \\
21 & 2011-09-13T22:10 & 2011-09-17T04:00 &335 & -12 & 3 & 184 & 5 & 3 & FL & 22:00 & 3 & 10 \\
22 & 2011-10-02T01:00 & 2011-10-05T08:00 &137 & 38 & 3 & 172 & 80 & 2 & FL(M3.9) & 0:37 & 14 & -3 \\
23 & 2011-10-22T00:05 & 2011-10-24T19:00 &291 & 40 & 2 & 286 & 45 & 2 & FE & 1:30 & 28 & 25 \\
24 & 2012-02-24T02:25 & 2012-02-26T22:00 &295 & 5 & 3 & 308 & 46 & 1 & FE & 2:00 & -32 & 28 \\
25 & 2012-06-08T03:20 & 2012-06-11T15:00 &90 & -35 & 3 & 73 & -65 & 1 & FL(C7.7) & 2:51 & 18 & -23 \\
26 & 2012-06-14T13:30 & 2012-06-16T10:00 &284 & -15 & 2 & 267 & -9 & 1 & FL(M1.9) & 12:52 & -9 & -20 \\
27 & 2012-07-03T20:45 & 2012-07-08T03:00 &161 & 36 & 3 & 185 & 76 & 3 & FL(C9.3) & 20:36 & 23 & 6 \\
28 & 2012-07-04T16:45 & 2012-07-08T03:00 &160 & -40 & 2 & 160 & -51 & 1 & FL(M1.8) & 16:33 & 33 & 6 \\
\hline
\end{tabular}
\end{center}
\tablecomments{
  $^a$MC solution from \citet{rpl11,rpl15}, with
  ($\phi_l$,$\theta_l$) the MC axis direction in HEE coordinates, and
  $Q_l$ a fit quality flag.  $^b$MC axis direction ($\phi_n$,$\theta_n$)
  in HEE coordinates and fit quality flag ($Q_n$)
  from an analysis newly presented here.  $^c$Type of surface
  activity: FL=flare (with GOES designation if available), FE=filament
  eruption.  $^d$Longitude and latitude of surface activity in HEE
  coordinates.}
\normalsize
\end{table}
     There are a total of 48 MCs identified in the time period
of interest.  The next step was to peruse the STEREO and SOHO/LASCO
imagery to identify CME candidates for these MCs.  Our approach was
to start far from the Sun in the HI2 field of view, in which
Earth is actually visible.  For each MC, we look at movies of HI2-A
and HI2-B images near the time of the MC to see if we can see a front
passing over Earth at about that time, which we then seek to track
backwards to HI1, then to COR2, and finally to COR1 to identify the CME
associated with that front and its initiation time at the Sun.
Only if we can identify a CME trackable continuously from COR1 to HI2
that appears to reach Earth at about the right time do we consider
ourselves to have a candidate CME for the MC.
It is of course simplistic to expect to see a front on top of
Earth at the exact time of the MC, as the apparent position of a
CME front depends on its projection into the 2-D image plane.  In
practice, the MC will be more likely observed some time after the
front is observed passing over Earth due to these projection effects,
and this time delay can become larger the farther the STEREO spacecraft
are from Earth.  These projection effects are naturally considered
when identifying candidate CMEs.

     For candidate CMEs, STEREO and SOHO/LASCO imagery are used
to perform a full 3-D kinematic and morphological reconstruction of
the CME, which will be described in detail in the next
section.  Only after such a reconstruction is deemed successful
at both adequately reproducing the appearance of the CME in images,
and in adequately reproducing the CME's arrival time at {\em Wind},
is the MC-CME connection finally deemed secure.  In some cases,
there are multiple CME candidates for an MC, and multiple CMEs must
therefore be modeled in order to establish the one most likely to be
responsible for the MC.  It is worth noting that there were a few
cases where a CME considered a candidate based on the initial
inspection of the images was ultimately deemed incorrect once
a detailed reconstruction was performed, the point being that
it is possible to be fooled into erroneous CME-ICME connections
if analysis is limited to visual inspection only.

     Nevertheless, most of our candidate CMEs were confirmed, and
we ultimately connect 31 out of the 48 {\em Wind} MCs with CMEs.
These CMEs are listed in Table~1.  There is only one MC event in all
of 2008 in the \citet{rpl11} list, and that is from
2008~December~16.  This is the first event listed in Table~1.  We emphasize
that we would have considered earlier 2008 MCs if there had been any.
In five cases there are two MCs identified very close in time, which
greatly complicates our interpretation.  For two of these cases
(\#20-21 and \#27-28 in Table~1) we associate the two MCs with two
distinct CMEs seen erupting from the Sun.  However, for three cases
(\#8, \#11, and \#18) we only see a single CME, and we end up
concluding that the one CME accounts for both MCs.  This is why there
are only 28 CMEs listed in Table~1 to account for the 31 MCs.
There is more discussion of the multi-MC cases in the Appendix.

     The CME start time listed in Table~1 corresponds to the time that
expansion is first perceived close to the Sun in COR1 images.  It
should be noted that for some of the slow CMEs in our list, the
expansion begins so gradually that there is ambiguity in defining a
clear start time.  The ICME arrival time at Earth is estimated from
{\em Wind} data time-shifted to the bow shock nose, extracted from
NASA/GSFC's OMNI database.  The {\em Wind} measurements of proton
density, velocity, and magnetic field are shown in Figure~1 for
our 28 CMEs.  The MC time intervals from \citet{rpl11,rpl15}
are indicated in the figure.
However, in most cases inspection of the {\em Wind} data leads us
to see the MC region as being just part of a larger ICME structure,
due to high velocities, densities, and/or fields extending beyond
the MC bounds.  We estimate the ICME arrival time
(to the nearest hour) based on the time of what we propose to be
the initial density, velocity, and/or field increase of the broader
ICME region.  This assessment is in some cases unambiguous
(e.g., events \#8 and \#18), but in other cases it is somewhat
subjective (e.g., events \#13 and \#25).
The assumed ICME arrival time is indicated
in Figure~1 and listed in Table~1.  Our events have travel times from
the Sun to Earth that range from 1.85 days (\#26) to 7.29 days (\#4).
\begin{figure}[p]
\plotfiddle{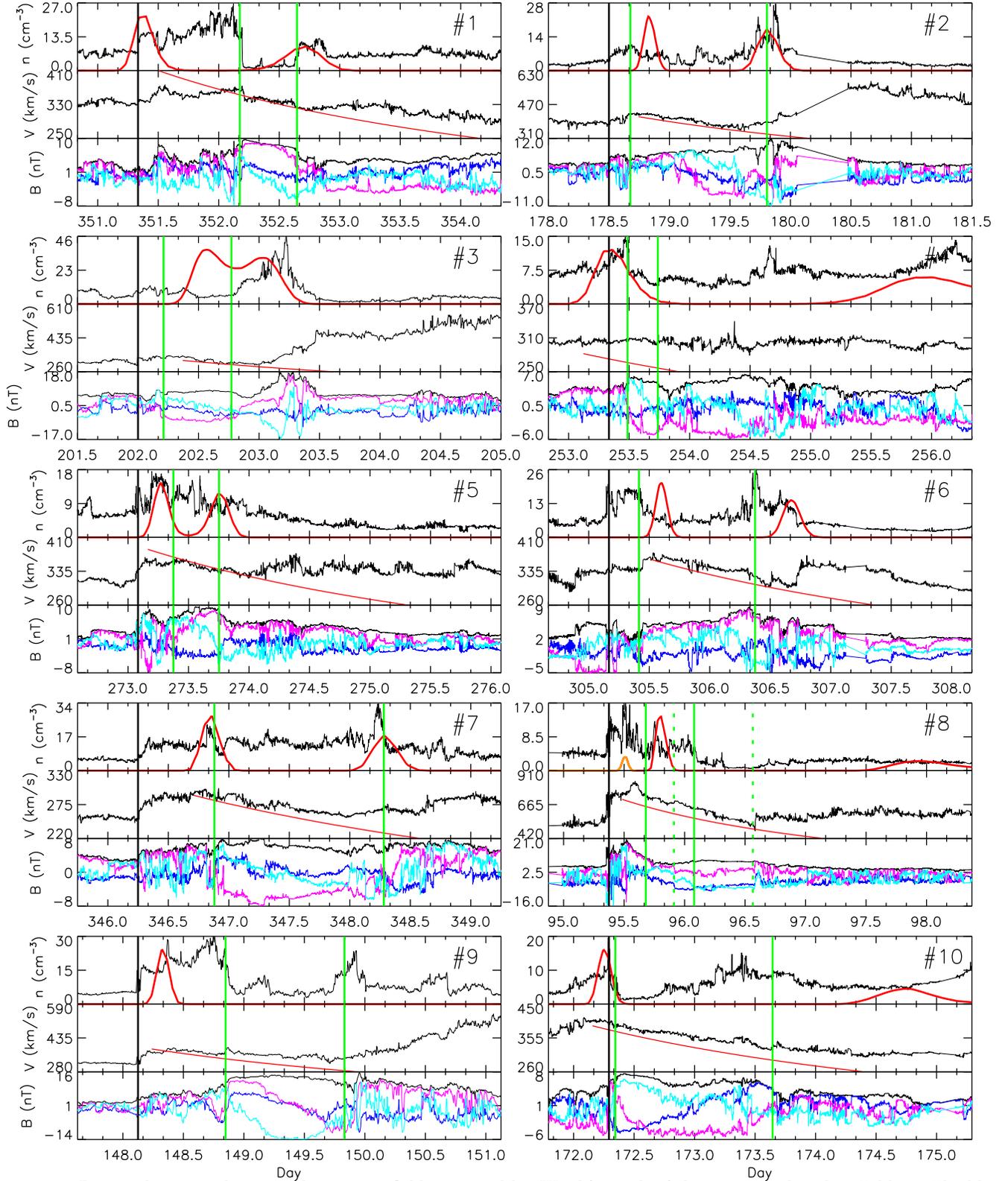}{8.2in}{0}{170}{170}{-520}{-630}
\caption{Proton density, velocity, and magnetic field measured by
  {\em Wind} for each of the 28 events listed in Table~1.  The blue, purple,
  and light blue curves in the magnetic field panel are $B_x$, $B_y$,
  and $B_z$, while the black line is $B_{tot}$.  Vertical black lines
  indicate the estimated ICME arrival time.  Vertical green lines indicate
  MC boundaries from \citet{rpl11,rpl15}.  In five panels there
  are two MCs identified, the second with dotted lines.  Red and orange
  lines show the density and velocity profiles predicted by the 3-D CME
  reconstruction described in Section~3, with red peaks in the density
  panel corresponding to the front and/or back of the FR, and orange peaks
  corresponding to a shock.  In two panels (\#20-21, \#27-28), there are
  two CMEs shown, the second using dotted instead of solid lines.}
\end{figure}
\setcounter{figure}{0}
\begin{figure}[t]
\plotfiddle{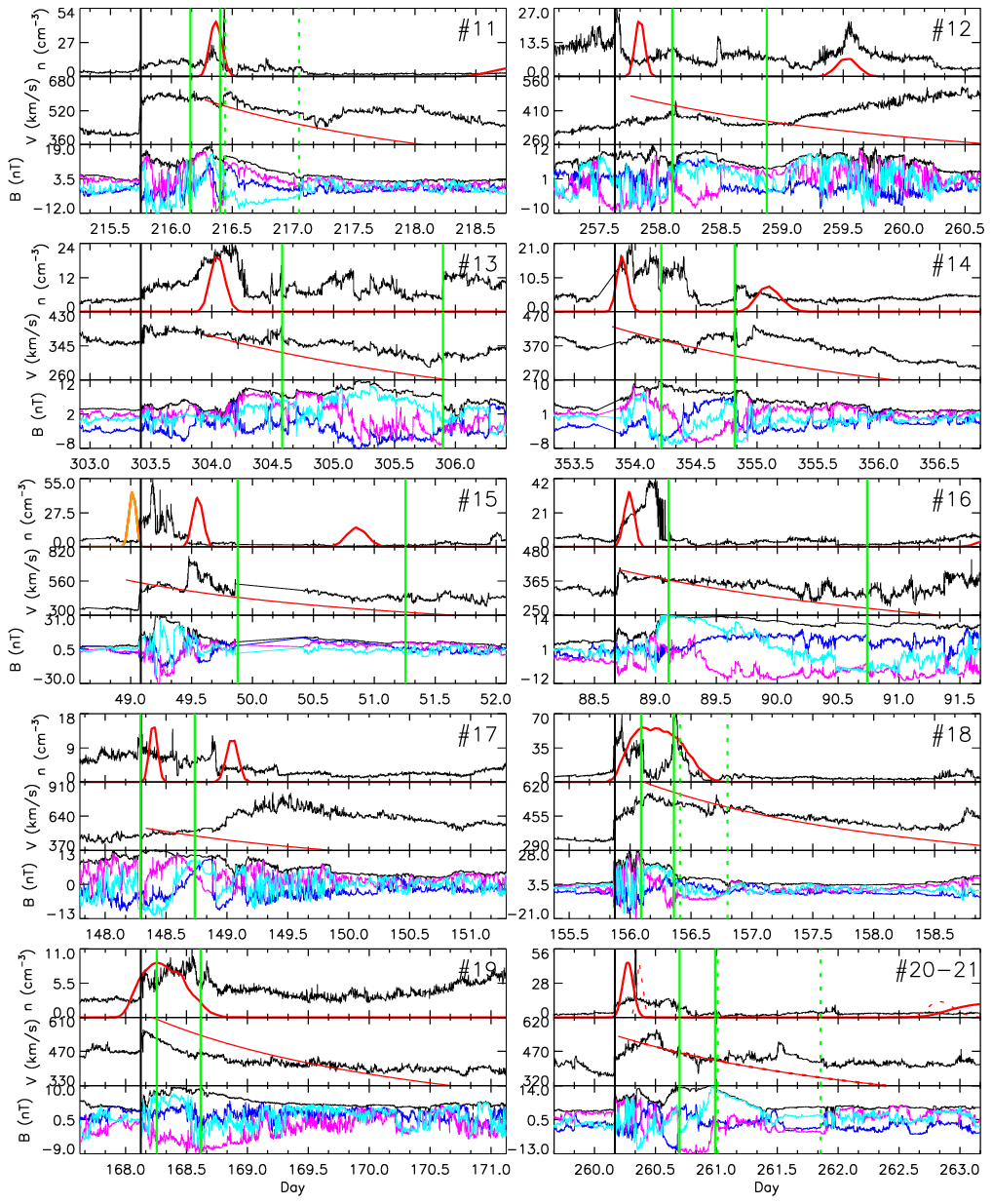}{8.7in}{0}{170}{170}{-520}{-620}
\caption{(continued)}
\end{figure}
\setcounter{figure}{0}
\begin{figure}[t]
\plotfiddle{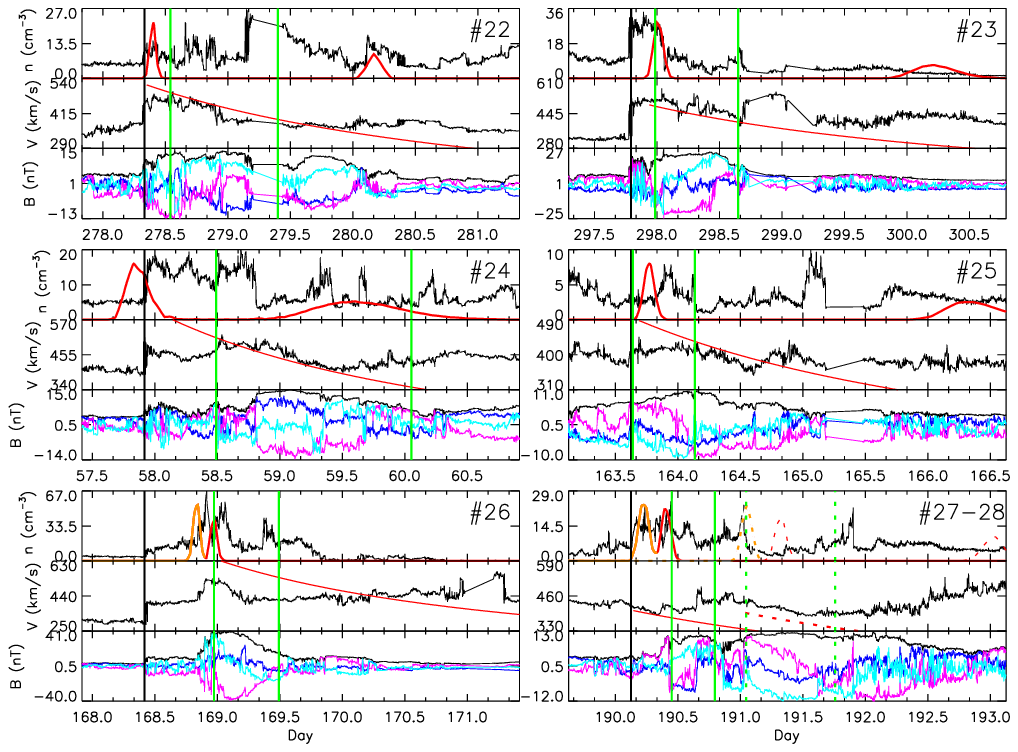}{5.0in}{0}{170}{170}{-520}{-860}
\caption{(continued)}
\end{figure}

     The $\phi_l$ and $\theta_l$ parameters in Table~1 indicate the
magnetic cloud orientations suggested by \citet{rpl11,rpl15}.
The analysis assumes the MC is a force-free FR with a cylindrical
geometry, with ($\phi_l$,$\theta_l$) indicating the attitude of the
axis of the cylinder, in the direction of the axial magnetic field.
Note that while \citet{rpl11,rpl15} use GSE coordinates, we
here convert to heliocentric Earth-ecliptic (HEE) coordinates, which
are more natural for remote sensing studies of CMEs propagating from
the Sun to Earth.  In HEE coordinates, the x-axis points from the Sun
towards Earth, the z-axis points from the Sun towards ecliptic north,
and the y-axis points to the right from Earth's perspective (while
facing the Sun), so $\tan\phi_l=y/x$ and $\tan\theta_l=z/\sqrt{x^2+y^2}$.
The $Q_l$ parameter in Table~1 is an MC fit quality
parameter from \citet{rpl11,rpl15}, with $Q_l$=1, 2, or 3
indicating good, fair, or poor.

     Partly to investigate systematic uncertainties in the MC parameters,
we have performed an independent analysis of the MC events using the
analysis approach described by \citet{tnc16},
relaxing the traditional force-free assumption used in the
\citet{rpl11,rpl15} fit.  Two examples of the
fits to the {\em Wind} magnetic field measurements are shown in
Figure~2.  The model assumes the same circular-cylindrical geometry
assumptions as in the \citet{rpl11,rpl15} fits.  However, it
does not assume any force distribution on the magnetic structure but
makes assumptions for the axial and poloidal current density
components.  We assume the $\tau=1$ case described in equation
(10) of \citet{tnc16} (i.e., the axial field is assumed to
go to zero at the outer edge of the FR).

     A reconstruction is based on a multiple regression technique
to infer the spacecraft trajectory using the Levenberg-Marquardt
algorithm. The parameters are initialized using minimum variance for
the orientation and magnetic field observations at the edge and center
for the current density components.  It is a semi-automatic iterative
exercise going from the local flux-rope coordinate system to the
spacecraft coordinate system to converge on the final parameters. The
quality of the fittings are based on the correlation coefficient
described by \citet{tnc16}.
The $\phi_n$ and $\theta_n$ parameters in Table~1 indicate the
resulting MC orientations, for the same boundaries as \citet{rpl11,rpl15},
and $Q_n$ is the fit quality indicator
from this new analysis, which can be compared with the analogous
$Q_l$ flag.
\begin{figure}[t]
\plotfiddle{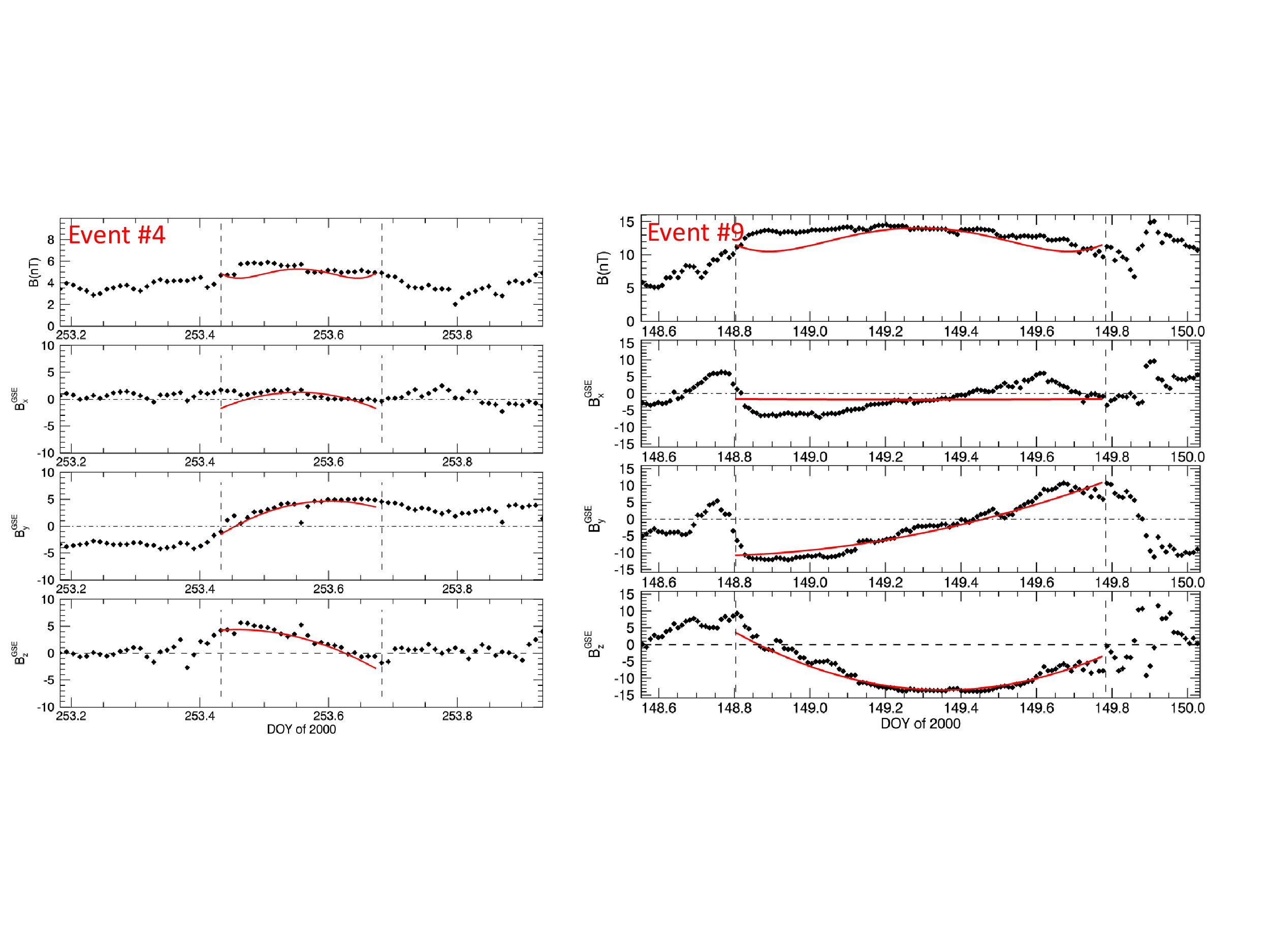}{2.9in}{0}{70}{70}{-245}{-90}
\caption{Two examples of magnetic cloud fits to {\em Wind} observations
  for events \#4 and \#9 in Table~1, using the \citet{tnc16}
  technique, leading to the ($\phi_n$,$\theta_n$) MC
  orientation measurements quoted in Table~1.}
\end{figure}

     Finally, for all our events, we have looked for solar surface
activity associated with each CME, particularly in EUV movies
from the EUVI imagers on STEREO, as well as images from the
Atmospheric Imaging Assembly (AIA) instrument on board the {\em Solar
Dynamics Observatory} (SDO).  We look for surface activity at the
right time and location to connect with the CMEs in our list.  Table~1
indicates which events are convincingly associated with significant
flares in the EUV, and which are associated with clear filament
eruptions.  In the case of flares, the GOES flare strength
designation is listed, if it has one.  The start time of the activity
is listed, which is naturally very close to the CME start time in all
instances.  The surface location of the activity in HEE coordinates is
also listed in Table~1.  For eight events in our sample, we find no
surface activity at all that can be definitively associated with the
CMEs.

     Our sample of events overlaps with other studies that have
connected CMEs observed by STEREO to ICMEs observed by spacecraft at
1~AU.  Relevant surveys that we are aware of include \citet{nl12b},
who studied ICMEs that impacted the STEREO spacecraft between 2008 and
early 2010; \citet{ekjk12,ekjk14}, who studied 1~AU travel times
of 31 events from 2008--2010 and searched for solar sources of twenty
2009 ICME events observed at Earth; \citet{cm14}, who
studied the kinematics of a sample of 24 CMEs from 2008--2012 that hit
{\em Wind} or one of the STEREO spacecraft; and \citet{bew16b},
who focused on STEREO observations of nine of the most geoeffective
CMEs from the first half of solar cycle 24.

     Nevertheless, we believe our list represents the first complete
STEREO survey of MC events in a multi-year time period.
We verify that the majority of MCs detected
at Earth are indeed ICMEs, as STEREO allows us to conclusively
connect 31 out of 48 (65\%) of the 2008--2012 {\em Wind} MCs with CMEs.
It is probable that many if not all of the remaining 17 (35\%)
will also be CMEs, but ones that did not allow for a truly
conclusive CME-MC connection.  Our conservative approach of
requiring CMEs to be trackable continuously from COR1 to HI2
will neglect cases of faint candidate CMEs that, for example, might
be tracked into HI1, but are too faint to be followed into HI2.
Obscuration by CIR fronts seen in HI2 \citep[see, e.g.,][]{bew10b}
or by other CME fronts could also play a role in
hiding candidate CMEs.  It is possible that some of the 17
unconnected MCs may not be ICMEs, but instead more quiescent
solar wind structures, perhaps connected with CIRs or SIRs.
It is worth noting that there is a tendency for the MCs successfully
connected to ICMEs to be longer than the unconnected MCs, with
the former having an average duration of 21.0~hr, and the latter
having an average duration of 13.6~hr.

\section{Reconstructing CME Kinematics and Morphology from STEREO Images}

     For each event in Table~1, we use STEREO and SOHO/LASCO images
to measure CME kinematics and morphology, operating within the
context of the flux rope paradigm for CME structure.  As an example,
Figure~3 shows a sequence of images following the 2012~February~24
CME (event \#24) from the Sun to near 1~AU.  The COR1 and COR2
images are displayed in a base-difference format, with an image from
before the CME subtracted from each CME image.  The LASCO/C3, HI1, and
HI2 images are shown in a running-difference format, with the
previous image subtracted from each image.  Movies showing the
sequence of STEREO and SOHO/LASCO images of all events are available
in the online version of this article.
\begin{figure}[t]
\plotfiddle{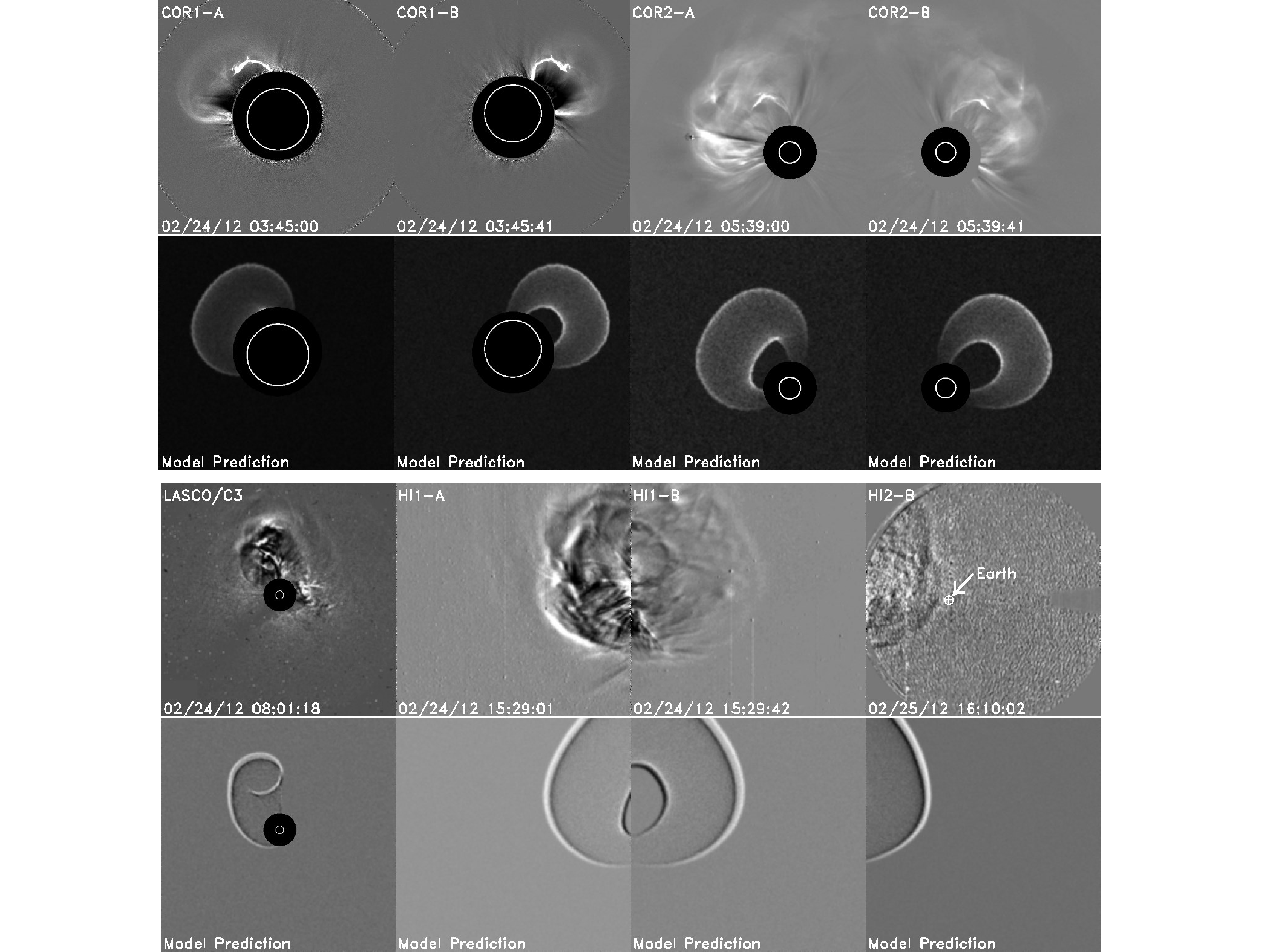}{5.5in}{0}{75}{75}{-270}{0}
\caption{A sequence of 8 STEREO and SOHO/LASCO images of a CME from
  2012~February~24 (event \#24), in chronological order.  Below
  each image is a synthetic image computed from the 3-D reconstruction
  of this event.}
\end{figure}

     Our techniques for 3-D CME reconstruction are already well
established in the literature
\citep{bew09a,bew09b,bew10a,bew11,bew12a,bew12b,bew16b},
though we briefly
describe them again here.  In our approach, the kinematic and
morphological analyses are performed separately, but there is
interdependence between the two.  Thus, in practice there is iteration
between the kinematic and morphological parts of the analysis to reach
a self-consistent final solution.

\subsection{Kinematic Analysis}

     We first focus on the kinematic analysis, which begins with
measurements of elongation angles of the CME leading edge from
Sun-center ($\epsilon$).  These measurements are made from a single
viewpoint, either STEREO-A or -B, depending on which image set
provides the clearest tracking of the CME front from COR1 to HI2.  The
elongation angles must be converted to actual distances from Sun
center ($r$), but this conversion requires geometric assumptions about
the shape of the CME leading edge.  If the CME is assumed to be
very narrow,
\begin{equation}
r=\frac{d\sin \epsilon}{\sin(\epsilon+\phi)},
\end{equation}
where $d$ is the distance from the observer to the Sun and $\phi$ is
the angle between the CME trajectory and the observer's line of sight
to the Sun \citep{swk07,nrs08}.  This can
be referred to as the ``Fixed-$\phi$'' approximation.
Another common assumption is that the
CME front can be approximated as a sphere centered halfway between the
Sun and the leading edge, leading to the so-called ``Harmonic Mean''
approximation \citep{nl09},
\begin{equation}
r=\frac{2d\sin \epsilon}{1+\sin(\epsilon+\phi)}.
\end{equation}
More complex assumptions about CME front geometry lead to
relations with more free parameters \citep[e.g.,][]{jad13}, but
we will here consider only the above two relations.

     There are three distinct constraints on the trajectory
direction quantified by $\phi$ in equations (1) and (2), which
can be called the ``stereoscopic constraint,'' the
``reasonable kinematics constraint,'' and the ``Earth arrival time
constraint.''  The first is by far the most important, which is
provided by comparing the angular extent of the CME from the
perspectives of STEREO-A and -B.  This assessment comes from the
morphological analysis described in Section~3.2, involving
reconstruction of the CME's shape as well as its trajectory direction.
The second constraint involves the assumption that at large distances
from the Sun the CME should trend towards a constant terminal speed.
In particular, there should be no implausible sudden accelerations or
decelerations far from the Sun.  The final constraint, which depends
on both the kinematic and morphological parts of the analysis, is that
the CME must hit Earth at the observed time (see Table~1).

     We experiment with both the ``Harmonic Mean'' and
``Fixed-$\phi$'' approximations, and use whichever is most
successful at finding a $\phi$ value that simultaneously addresses all
three constraints mentioned above.  A previous analysis of an event
from 2008~June~1 event provides an excellent example of a case where
the $\phi$ value inferred from the ``stereoscopic constraint'' yields
a far more reasonable kinematic profile (i.e., the ``reasonable
kinematics constraint'') when ``Harmonic Mean'' is assumed than
when ``Fixed-$\phi$'' is used \citep[see Figure~2 in][]{bew10a}.
Our survey now allows us to say that this is indeed the norm, and that
the ``Harmonic Mean'' approximation is usually far more successful in
this regard than ``Fixed-$\phi$.''  There are only four cases (events
\#5, \#13, \#17, and \#25) where we end up preferring
``Fixed-$\phi$.''

     Once the CME leading edge distances ($r$) are established,
we can study CME kinematics, starting with plots of $r$ versus
time.  The top panel of Figure~4(a) shows such a plot
for event \#24.  From such measurements we seek to
infer velocity and acceleration profiles.  For that purpose,
we use a simple kinematic model that we have used many
times in the past \citep{bew09a,bew09b,bew10a,bew11,bew12a,bew12b,bew16b},
which divides the CME travel time into $2-4$ phases
of constant velocity and/or acceleration (or deceleration).  Decisions
on the type and number of phases are made after inspecting
velocity-versus-time profiles computed point-by-point.  The
kinematic model for event \#24 in Figure~4(a) is particularly simple,
with only two phases: a constant acceleration phase, followed by a
constant velocity phase.
\begin{figure}[t]
\plotfiddle{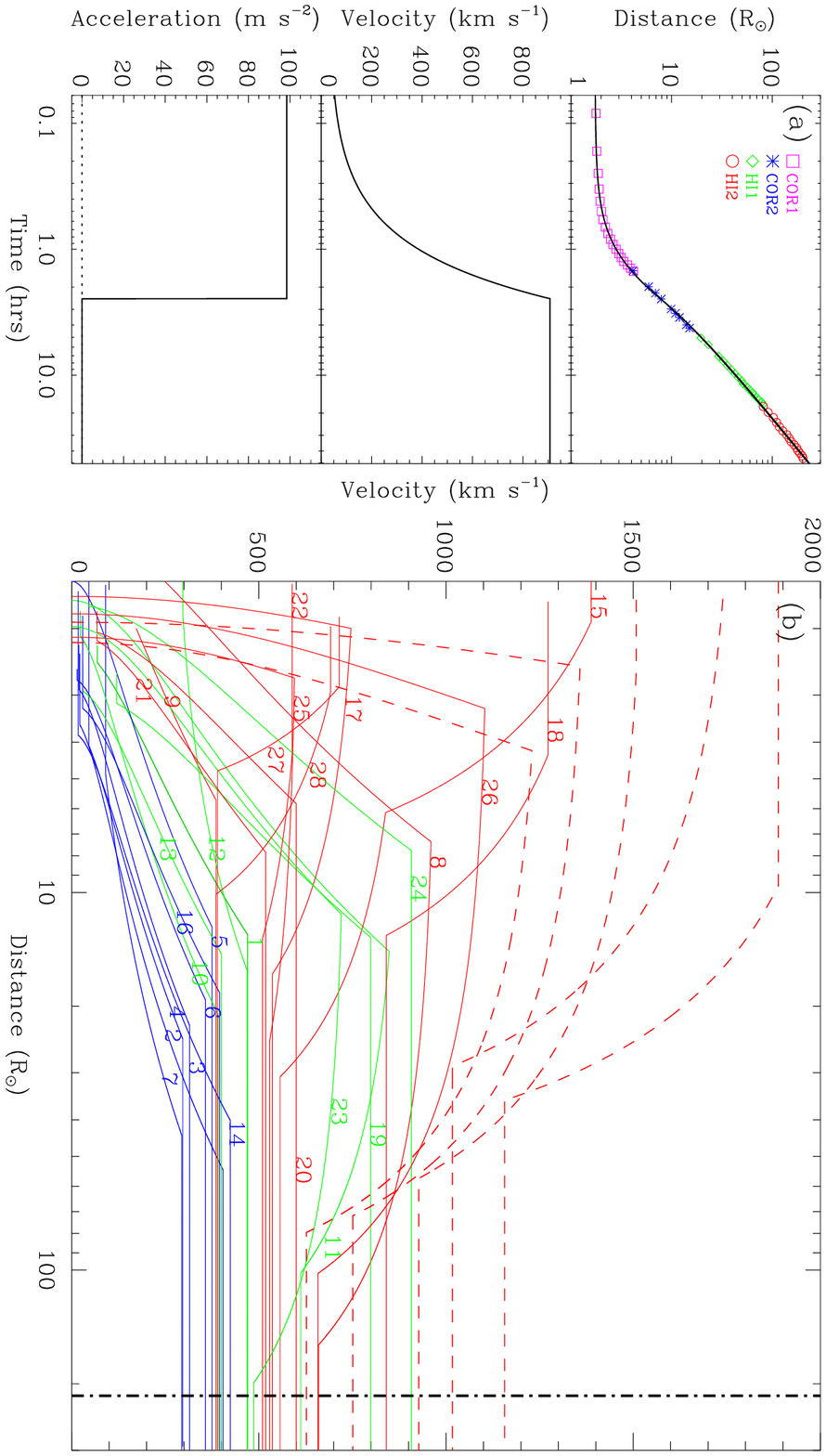}{3.9in}{90}{75}{75}{295}{-80}
\caption{(a) Kinematic model for event \#24 in Table~1, with
  the top panel showing distance versus time measurements for the
  leading edge of the CME, from STEREO-B for this particular
  event. These measurements are fitted with a simple two-phase
  kinematic model, yielding the solid line fit to the data, and the
  velocity and acceleration profiles shown in the lower panels.
  (b) Velocity versus distance profiles for the leading edges
  of all 28 CMEs listed in Table~1, with line color indicating
  solar surface activity associated with the events.  Red, green,
  and blue indicates association with flares, filament eruptions
  (but no flares), and no surface activity at all, respectively.
  The red dashed lines are additional kinematic profiles from
  \citet{bew16b}, for five geoeffective events not perceived
  as MCs at Earth.  The dot-dashed line indicates 1~AU.}
\end{figure}

     Assuming 1\%, 2\%, and 3\% uncertainties
for the distance measurements from COR1/COR2, HI1, and HI2,
respectively \citep[e.g.,][]{bew09a}, we use a $\chi^2$
minimization routine to find the best fit to the distance
measurements \citep{prb92}.  For event \#24, the
top panel of Figure~4(a) shows the best fit to the data, and
the panels below show the corresponding velocity and acceleration
profiles.  For this two-phase fit there are only three free
parameters: a starting height, the acceleration for
the first phase, and a time for the end of that phase.

     Figure~4(b) shows the velocity versus distance profiles of all
28 CMEs in our sample.  From these curves we can infer a peak
velocity ($V_{max}$) and a terminal speed far from the Sun ($V_{term}$).
These values are listed in Table~2, along with a number of
morphological parameters that will be described in Section~3.2.
Our MC-selected CME sample includes only three events that reach
$V_{max}>1000$ km~s$^{-1}$, and none with $V_{max}>1500$ km~s$^{-1}$.
In order to increase the number of fast CMEs in our sample, we have in
Figure~4(b) added kinematic profiles for five additional events from
\citet{bew16b}, who studied nine geoeffective CMEs, performing
kinematic and morphological reconstructions for eight of them based on
STEREO images.  Three of these eight are already in our sample (events
\#8, \#18, and \#23), but five are not.  These events, with initiation
dates of 2011~August~3, 2011~September~24, 2012~March~7,
2013~October~29, and 2014~February~25, all have $V_{max}>1000$
km~s$^{-1}$ but were not perceived as MCs at Earth, despite being very
geoeffective.  These are the five events whose kinematic profiles are
shown as dashed lines in Figure~4(b).

\begin{table}[t]
\scriptsize
\begin{center}
\caption{CME Measurements}
\begin{tabular}{cccccccccccccc} \hline \hline
ID & $\lambda_s$$^a$ & $\beta_s$$^a$ & $\gamma_s$$^b$ & FWHM$_s$$^c$ & $\Lambda_s$$^d$ & $\eta_s$$^e$ &
  $\alpha_s$$^f$ & Q$_s$$^g$ & V$_{max}$$^h$ & V$_{term}$$^h$ & $D_{st}$$^i$ & $\phi_s$$^j$ & $\theta_s$$^j$ \\
   &   (deg)     &   (deg)   &  (deg)     & (deg) &   &          &
             &    &(km~s$^{-1}$)&(km~s$^{-1}$)&  (nT) & (deg)   &   (deg)    \\
\hline
1 & 9 & 10 & -60 & 77.4 & 0.25 & 1.1 & 2.5 & 4 & 470.4 & 470.4 & -15 & 102.9 & -51.1 \\
2 & 13 & -10 & -20 & 70.8 & 0.10 & 2.31 & 8.0 & 4 & 404.4 & 404.4 & -28 & 70.1 & -20.8 \\
3 & -9 & -6 & -8 & 62.0 & 0.07 & 1.0 & 8.0 & 2 & 315.0 & 315.0 & -83 & 100.0 & -5.6 \\
4 & -25 & -5 & -5 & 76.3 & 0.18 & 1.6 & 6.0 & 3 & 296.9 & 296.9 & -14 & 117.7 & 0.9 \\
5 & 1 & -2 & -10 & 41.5 & 0.06 & 1.5 & 4.0 & 1 & 374.7 & 374.7 &  -9 & 87.1 & -10.0 \\
6 & 8 & 1 & -4 & 41.5 & 0.10 & 1.0 & 4.0 & 2 & 357.6 & 357.6 & -10 & 73.2 & -3.1 \\
7 & 9 & -5 & -40 & 52.1 & 0.12 & 1.5 & 3.0 & 3 & 294.2 & 294.2 &  -2 & 68.9 & -33.9 \\
8 & 3 & -16 & -80 & 59.9 & 0.21 & 1.6 & 4.0 & 3 & 959.9 & 659.9 & -81 & 12.4 & -43.6 \\
9 & 12 & 3 & 50 & 76.3 & 0.31 & 1.0 & 6.0 & 3 & 385.9 & 385.9 & -80 & 71.0 & 40.3 \\
10 & -11 & 10 & -25 & 64.5 & 0.17 & 1.6 & 4.0 & 3 & 391.0 & 391.0 & -11 & 144.6 & -15.7 \\
11 & -6 & 20 & -40 & 80.5 & 0.30 & 1.3 & 3.0 & 2 & 848.1 & 612.0 & -74 & 128.4 & -35.9 \\
12 & -8 & 15 & -35 & 104.3 & 0.20 & 1.7 & 5.0 & 4 & 468.4 & 468.4 & -25 & 93.9 & -34.3 \\
13 & 9 & -30 & -65 & 111.9 & 0.22 & 1.4 & 3.5 & 4 & 399.5 & 399.5 & -11 & 20.0 & -39.6 \\
14 & 10 & -10 & -25 & 73.7 & 0.12 & 2.0 & 4.0 & 3 & 423.8 & 423.8 & -35 & 61.8 & -23.4 \\
15 & -8 & -10 & 45 & 92.0 & 0.15 & 2.0 & 2.5 & 3 & 1387.8 & 556.8 & -31 & 114.9 & 38.3 \\
16 & -11 & 5 & -25 & 82.2 & 0.25 & 1.4 & 4.5 & 3 & 395.2 & 395.2 &  -4 & 101.6 & -23.7 \\
17 & 1 & -5 & -80 & 62.8 & 0.10 & 1.5 & 6.0 & 4 & 746.2 & 536.3 & -80 & 72.7 & -66.9 \\
18 & -25 & -5 & -40 & 83.0 & 0.31 & 1.1 & 4.0 & 2 & 1272.0 & 840.3 & -45 & 108.0 & -27.9 \\
19 & -39 & -5 & 35 & 119.8 & 0.18 & 1.6 & 4.0 & 4 & 798.9 & 798.9 & -14 & 122.6 & 17.2 \\
20 & 23 & 10 & 35 & 85.3 & 0.23 & 1.4 & 3.0 & 3 & 599.6 & 599.6 & -69 & 41.7 & 19.0 \\
21 & -2 & 15 & -75 & 73.7 & 0.25 & 1.3 & 4.0 & 3 & 518.2 & 518.2 & -69 & 142.6 & 13.5 \\
22 & -1 & -3 & -50 & 59.0 & 0.21 & 1.0 & 5.0 & 3 & 589.1 & 510.0 & -43 & 89.8 & -43.7 \\
23 & -5 & 25 & -60 & 98.7 & 0.19 & 1.5 & 6.0 & 3 & 720.3 & 486.8 &-147 & 147.7 & 12.8 \\
24 & -26 & 35 & -73 & 118.4 & 0.29 & 1.25 & 3.0 & 5 & 907.6 & 907.6 & -48 & 151.5 & -35.7 \\
25 & 2 & -20 & -80 & 75.8 & 0.19 & 1.6 & 3.0 & 3 & 595.0 & 528.5 & -51 & 9.9 & -43.1 \\
26 & -3 & -20 & -10 & 113.7 & 0.43 & 1.6 & 3.0 & 2 & 1103.5 & 657.7 & -71 & 90.0 & -9.7 \\
27 & -1 & -10 & -85 & 66.3 & 0.25 & 1.0 & 3.0 & 2 & 715.0 & 389.7 & -68 & 9.7 & -53.1 \\
28 & -1 & 10 & 30 & 83.0 & 0.19 & 1.3 & 4.0 & 2 & 691.8 & 385.6 & -68 & 89.5 & 28.6 \\
\hline
\end{tabular}
\end{center}
\tablecomments{$^a$Longitude and latitude of central FR trajectory, in
  HEE coordinates.  $^b$Tilt angle of FR ($0^{\circ}$ is an E-W
  orientation).  $^c$FR angular full width at half-maximum.  $^d$FR
  aspect ratio (apex radius divided by distance from Sun).
  $^e$Ellipticity of FR cross section.  $^f$Parameter defining shape
  of FR leading edge.  $^g$Quality flag ($1-5$) assessing the evidence
  for an FR shape (1=no evidence, 5=strong evidence).  $^h$Peak and
  terminal CME speeds.  $^i$Minimum $D_{st}$ value, as a measure of
  geoeffectiveness.  $^j$FR axis direction seen by Earth, in HEE
  coordinates.}
\normalsize
\end{table}

     Figure~4(b) divides the CMEs into three categories, based on what
solar surface activity is associated with the events.  We will refer
to events associated with flares as ``group~1,'' events associated
with filament eruptions but no flares ``group~2,'' and events
with no surface activity ``group~3.''  If the five extra geoeffective
events are included, groups~1, 2 and 3 include 17, 8, and 8 events,
respectively.  The group~3 events are all slow, with peak speeds
of $V_{max}=294.2-423.8$ km~s$^{-1}$.  The group~1 CMEs are all
faster than this, with one exception (event \#9).  Group~2 is
intermediate, with half being slow ($V_{max}<500$ km~s$^{-1}$) and
half relatively fast, though none reach $V_{max}>1000$ km~s$^{-1}$.

     Another very clear distinction in the kinematic behavior of these
three groups regards the height at which the peak velocity is reached.
This is shown explicitly in Figure~5(a).
We compute the average and standard deviation of this height for
all three categories, and find $R=3.2\pm 2.2$~R$_{\odot}$,
$R=13.9\pm 3.8$~R$_{\odot}$, and $R=29.4\pm 14.8$~R$_{\odot}$
for groups 1, 2, and 3, respectively.  Note that for group~1,
EUVI images of the CME fronts closer to the solar surface
would be required to precisely assess the height when peak velocity
is reached, as about half of these events are already at or near
their peak speeds when they enter the COR1 field of view at
$R\approx 1.5$~R$_{\odot}$.
\begin{figure}[t]
\plotfiddle{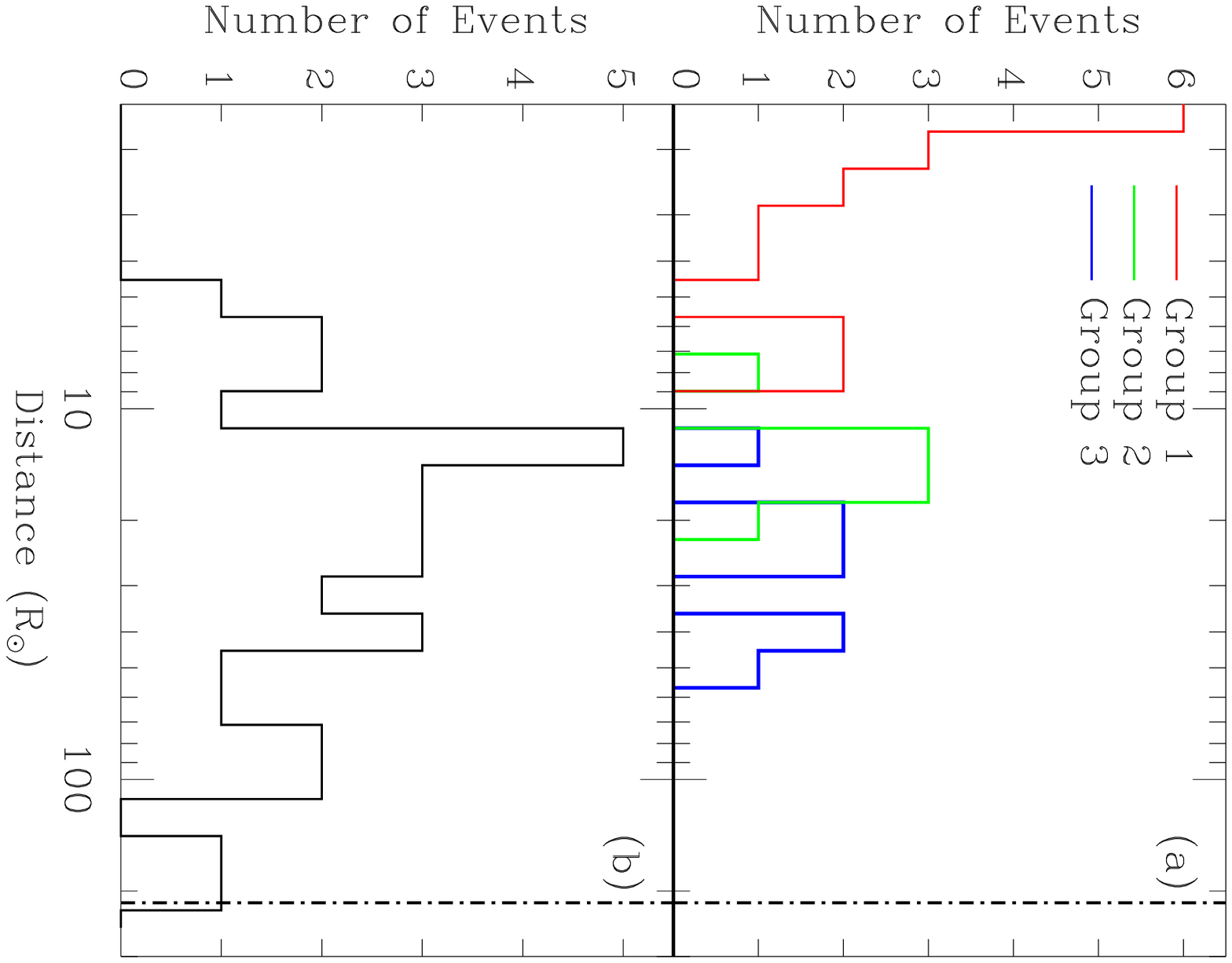}{3.6in}{90}{55}{55}{230}{-35}
\caption{(a) Distribution of heights at which CMEs first reach their
  peak velocities for group~1 (flare-associated CMEs),
  group~2 (filament eruption associated CMEs), and group~3
  (no surface activity). (b) Distribution of heights at which CMEs
  reach their final terminal speeds.  The vertical dot-dashed line
  indicates 1~AU.}
\end{figure}

     As shown in Figure~4(b), all but two events that reach
$V_{max}>600$ km~s$^{-1}$ subsequently decelerate, the example in
Figure~4(a) being one of the two exceptions.  The deceleration is
presumably due to drag forces induced by collision with the slower
ambient solar wind.  For a few CMEs (e.g., events \#15 and \#27)
deceleration seems to happen very early at $R<10$~R$_{\odot}$, but in
most cases the deceleration happens mostly further out.  Figure~5(b)
shows the distributions of heights at which CMEs reach their final
terminal speeds.  For fast events, this would generally be where
deceleration ceases, while for slow events this would be where
acceleration to the terminal speed stops.  The CMEs reach their final
speeds at heights ranging from $R=4.8-198.7$~R$_{\odot}$.  Half of our
events have reached their final speeds by $R=20$~R$_{\odot}$, while
80\% have reached it by $R=0.3$~AU (64.5~R$_{\odot}$).  Finally, it
should be emphasized that contrary to the velocity profiles in
Figure~4(b), it is likely that many CMEs still exhibit some degree of
deceleration at 1~AU.  Our simple, multi-phase kinematic modeling
approach neglects this by invariably assuming a final phase of
constant velocity.  Quantifying residual CME deceleration at 1~AU
would ideally require observational tracking of the CMEs well beyond
1~AU.

\subsection{Morphological Analysis}

     For the morphological part of our analysis, we use a
parametrized prescription for generating 3-D FR shapes.  For each such
shape, a 3-D density distribution is created with density placed on
the boundary of the FR shape, and these density cubes are then used to
compute synthetic images for comparison with real ones.  Parameters
are varied until agreement with the data is maximized.  Probably the
most widely used prescription for FR shape generation, and comparison
with STEREO imagery, is that of \citet{afrt06,at09}, for
which there is a user friendly IDL tool.  We instead use a procedure
first described by \citet{bew09a}, and which has been
subsequently used in many published analyses
\citep{bew10a,bew11,bew12a,bew12b,bew16b}.

     The inner and outer edges of the FR are defined in
polar coordinates with the equation
\begin{equation}
r(\theta)=r_{max} \exp \left( - \frac{1}{2} \left|
  \frac{\theta}{\sigma} \right| ^{\alpha} \right).
\end{equation}
This is a polar Gaussian if $\alpha=2$.  Higher $\alpha$ values create
loops with flatter, less rounded leading edges.  The $\sigma$
parameter determines the widths of the loops, which can also be
described by a full-width-at-half-maximum,
\begin{equation}
FWHM_s=2\left( 1.386 \right)^{1/\alpha} \sigma.
\end{equation}
As described by \citet{bew09a}, the two 2-D loops defining the
inner and outer edges of the FR are used to create a 3-D FR by first
creating a sequence of lines normal to the inner edge, extending to
the outer edge.  The midpoints of these lines define the central axis
of the FR.  A 3-D FR can be created by assuming circular cross
sections for the flux rope in all the planes defined by the normal
lines.  By stretching the FR in the direction perpendicular to the FR
creation plane, an FR can be created with an arbitrary ellipticity,
with the parameter $\eta_s$ describing the ratio of FR channel width
perpendicular to the FR plane to its width within the FR plane.

     The two main advantages of this FR shape prescription to
that of \citet{afrt06,at09} are the ability to assume
elliptical FR channels, and the ability (by increasing $\alpha$) to
assume FRs with flattened fronts.  Note that \citet{mj13,mj14}
have developed prescriptions for describing CME front shapes
that could easily replace equation (3), and then be used to create 3-D
FR shapes.  With straighter, more cone-like sides, these prescriptions
may be better at describing the shapes of the trailing parts of CME
fronts \citet{mj15}.

     There are three parameters describing how to orient the
3-D FRs.  A longitude and latitude in HEE coordinates
($\lambda_s$,$\beta_s$) describe the trajectory direction of the
center of the FR.  A tilt angle $\gamma_s$ describes how the FR axis is
oriented relative to the ecliptic plane, with $\gamma_s=0^{\circ}$
corresponding to an east-west orientation, parallel to
the ecliptic, and $\gamma_s=\pm
90^{\circ}$ corresponding to a north-south orientation, perpendicular
to the ecliptic.  A density cube containing a 3-D FR described by the
above parameters is created, placing density only on the surface of
the FR, assuming a Gaussian density profile normal to the surface.
\citet{bew09a} list a number of parameters describing how
density is placed on the FR surface, but these parameters are not of
much interest for our purposes, so we do not discuss them here.

     In order to confront the assumed FR shape with data, we first
expand it to the appropriate size using the kinematic model
described in Section~3.1.  We assume that the same CME structure
applies at all times during the CME's expansion outwards.  In other
words, we assume purely radial and self-similar expansion.  In
certain past analyses we have relaxed these assumptions and allowed
for time-dependent evolution of CME morphology \citet{bew10a,bew12a,bew12b},
but this is too time consuming for a survey of many
events.  Our impression is that self-similar expansion is a decent,
albeit not perfect, approximation for CMEs expanding into
the interplanetary medium.

     Synthetic images are then generated from the density cube
using a white-light rendering routine to perform the
necessary calculations of Thomson scattering within the density cube
\citep{deb66,afrt06}.  In this way, we can
generate synthetic images of the CME at any time and from any
perspective that we want, for comparison with images from STEREO-A,
STEREO-B, and SOHO/LASCO.  Parameters are adjusted to maximize
agreement between the real and synthetic images.  This is done
by trial-and-error, and subjective judgment is used to decide
what is the best fit.  Figure~3 shows synthetic images computed
in this manner for the 2012~February~24 CME (event \#24).
Considering the number of images we have to work with, we in
practice compare real and synthetic {\em movies} of the images.
Movies comparing the real and synthetic STEREO and SOHO/LASCO
images are available in the online version of this article, for
all the events in our sample.

     A secondary consideration in identifying the best fit parameters
is the agreement between the observed and predicted Earth arrival
time.  This is an issue for the morphological as well as the kinematic
part of the analysis, as changing the CME's shape or trajectory
direction can affect the Earth arrival time as much as changing the
kinematics of the CME leading edge.  The density and velocity profiles
predicted by our best models are compared with the {\em Wind} data in
Figure~1.  Since we only place density on the surfaces of our
FRs, the very simplistic model prediction is for two density peaks,
when {\em Wind} enters and exits the FR.  (In some cases the exit
peak is beyond the time range displayed in Figure~1.)  The predicted
velocity profile is always one of decreasing velocity, which is a
characteristic of self-similar expansion, where velocity within
the CME must be proportional to distance from the Sun for the CME
to stay the same shape.  These velocity decreases are similar to
those that are generally observed within the MC time range, which
will be discussed further in Section~5.3.

\begin{figure}[p]
\plotfiddle{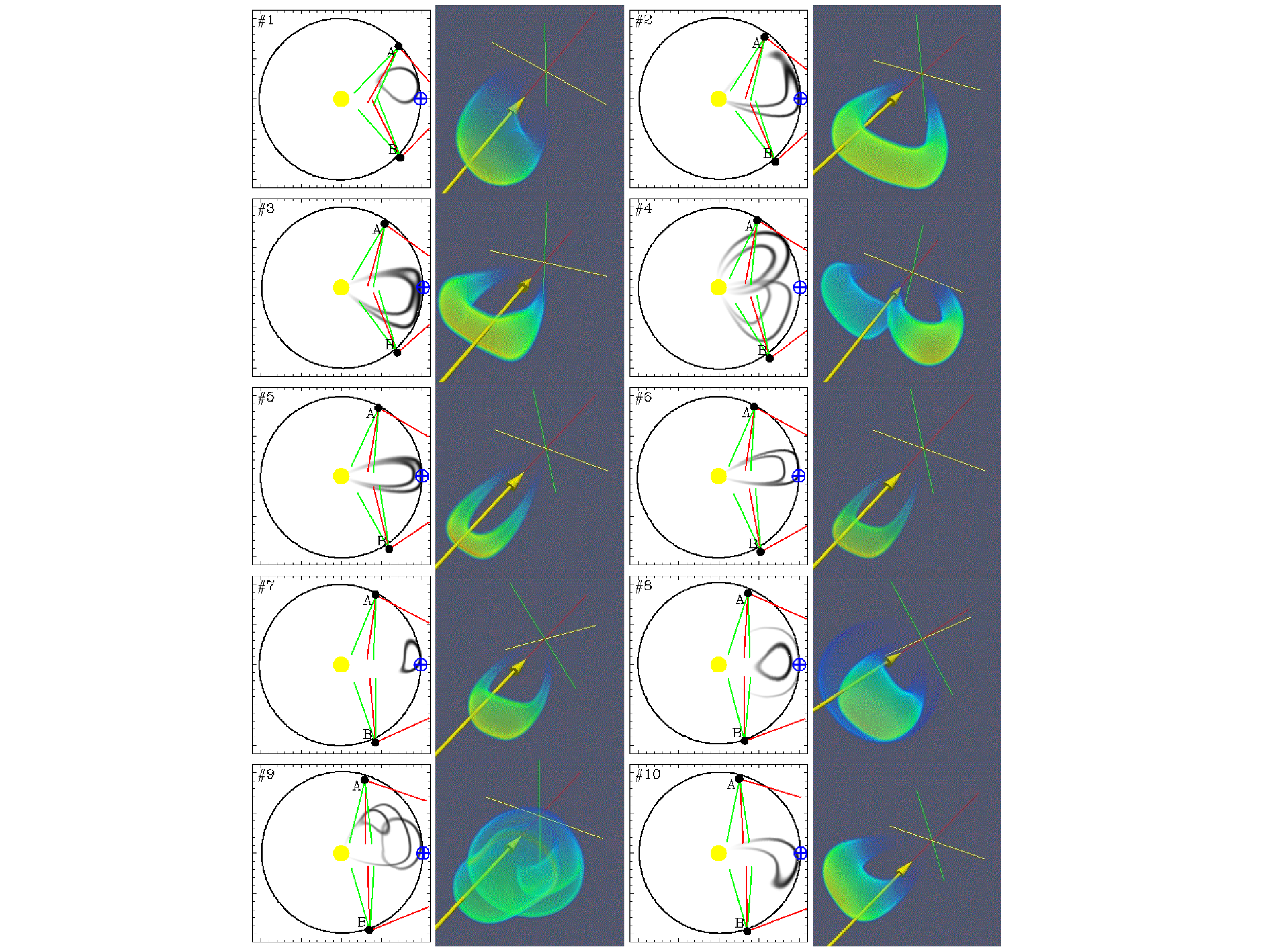}{7.9in}{0}{110}{110}{-390}{-5}
\caption{Two illustrations of the CME reconstruction are provided for
  each of the 28 events listed in Table~1.  The left panels are
  ecliptic plane maps showing the positions of Earth, STEREO-A, and
  STEREO-B in HEE coordinates, with the greyscale image being the
  ecliptic plane slice through the 3-D CME reconstruction shown in the
  right panel.  The green and red lines show explicitly the HI1 and
  HI2 fields of view.  In the right panel, the x-, y-, and z-axes are
  in red, yellow, and green, respectively.  The xy-plane is the
  ecliptic, with the x-axis defined by the Sun-Earth line, and the
  positive z-axis direction being ecliptic north (always up in each
  figure).  The yellow arrow indicates the Earth's path through the
  CME structure.  Note that many of the reconstructions actually
  include multiple CMEs (see text for details).}
\end{figure}
\setcounter{figure}{5}
\begin{figure}[t]
\plotfiddle{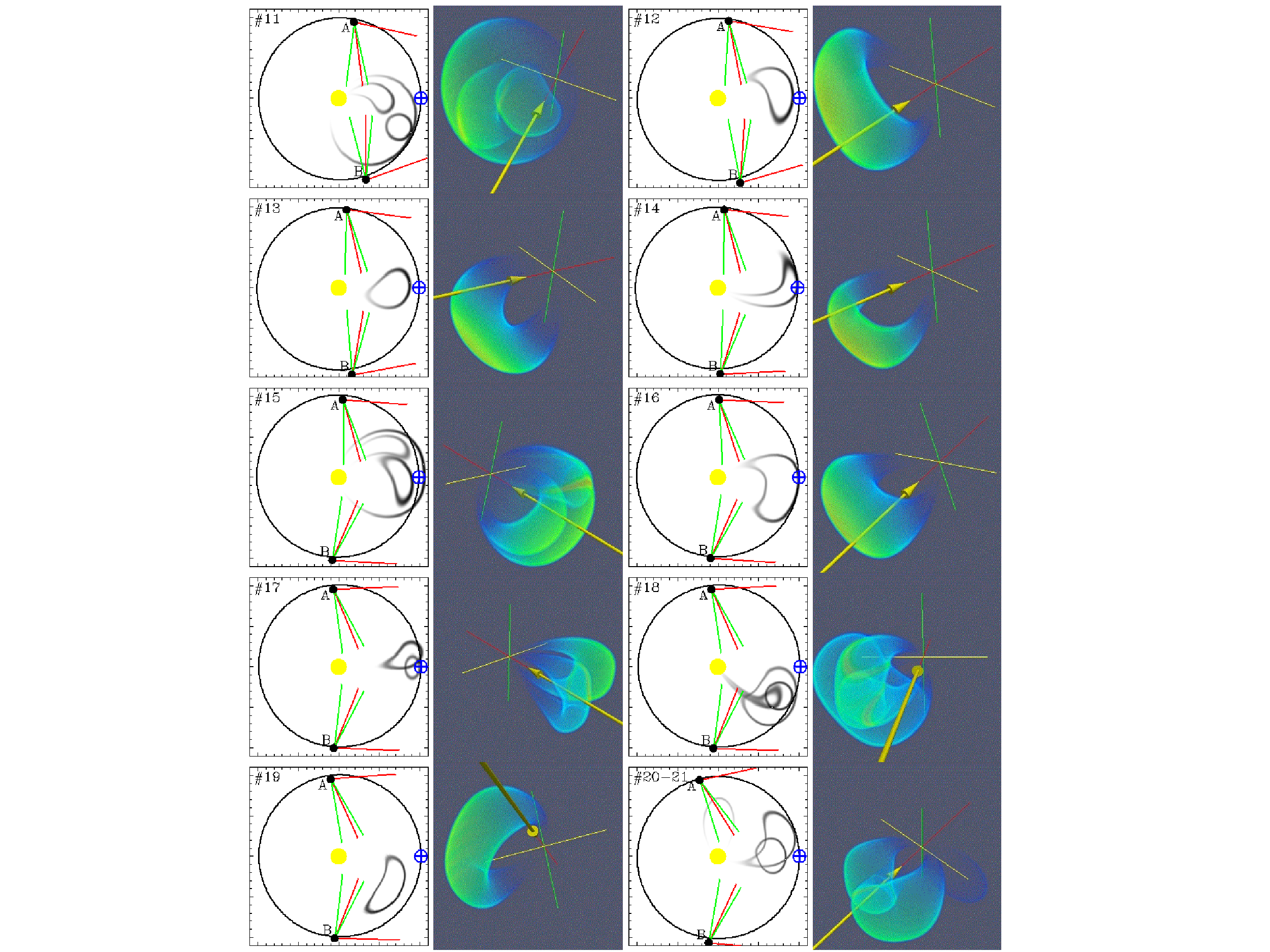}{8.7in}{0}{110}{110}{-390}{0}
\caption{(continued)}
\end{figure}
\setcounter{figure}{5}
\begin{figure}[t]
\plotfiddle{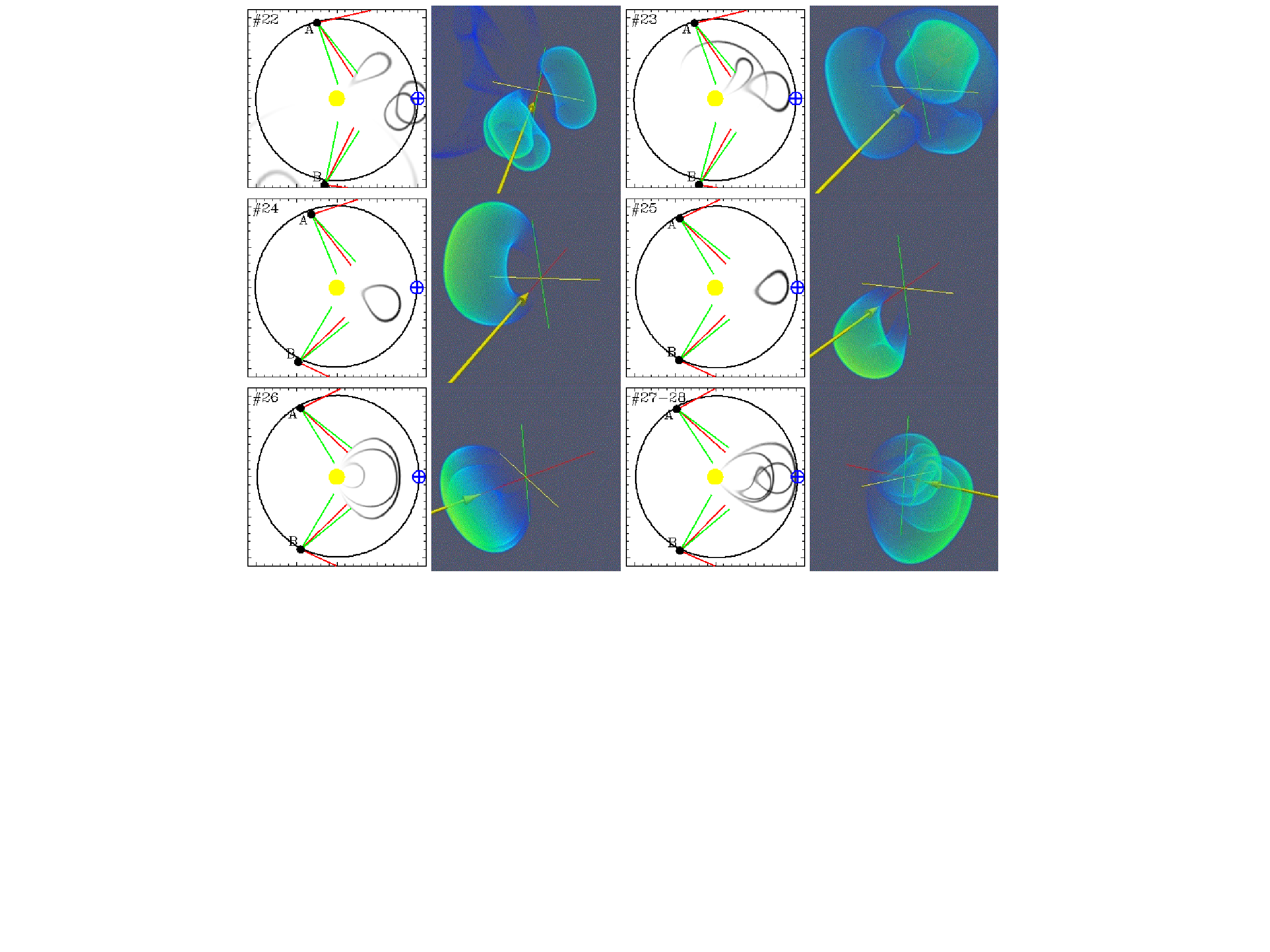}{4.9in}{0}{110}{110}{-390}{-240}
\caption{(continued)}
\end{figure}
     Figure~6 shows two different visualizations illustrating the 3-D
CME reconstructions, the right one being a picture of the 3-D structure
in HEE coordinates, with an arrow indicating the Earth's path
through it as the CME expands outwards, and the left one being a
map in the ecliptic plane displaying a slice through the
3-D density cube in that plane.  The ecliptic plane maps
indicate the STEREO-A, STEREO-B, and SOHO/LASCO (near Earth)
viewing geometry for each event, with the HI1 and HI2 fields of
view explicitly displayed for both STEREO-A and -B.  Columns
2--8 in Table~2 list the best-fit parameters of the FR structures.
Note that we are using ``s'' subscripts for all morphological
parameters derived from empirical analysis of STEREO and SOHO/LASCO
images.  The $\Lambda_s$ parameter is an aspect ratio dividing the
FR apex radius by its distance from the Sun.  A larger $\Lambda_s$
corresponds to a fatter FR.  The $\alpha_s$ parameter in Table~2,
from equation~(3), is the one for the leading edge of the FR and
not the trailing edge, which can be different.

     The images of some of our faster events show a front out ahead
of what we interpret as the FR ejecta, which is naturally interpreted
as a shock.  If this shock is sufficiently bright and trackable for
a long distance from the Sun, we model it as well as the FR, as
we have done for past events \citep[e.g.,][]{bew09a}.  To model
a shock front shape, equation~(3) is used to create a 2-D loop and
then that loop is rotated about its center to create a 3-D lobular
front.  As in the FR modeling, the various parameters are adjusted
to maximize agreement with the images.  We ended up including shocks
in the 3-D reconstructions of only five of the events in the Table~1
sample, specifically events \#8, \#15, \#26, \#27, and \#28 (see
Figure~6).  It is important to note that in the kinematic analysis in
Section~3.1 we always seek to follow what we perceive as the top of
the FR ejecta, and not the shock, although the two can be difficult to
distinguish sometimes, particularly in HI1/HI2.

     It is not uncommon for multiple CMEs to be visible
simultaneously in the STEREO and SOHO/LASCO images.  In some cases,
the other CME fronts are distinguishable from the one of interest to
an extent that we simply ignore the other CMEs, but in some instances
we found it desirable to reconstruct multiple CMEs to ensure that we
are correctly identifying the one that hits Earth and is responsible
for the MC.  Thus, in Figure~6 multiple CME structures are
sometimes seen.  In most cases the extra CMEs that are modeled
are not Earth-directed, and are therefore distinguishable from
the MC-related events of interest in Figure~6.  But there are some
exceptions (e.g., event \#22), and there are also the two cases of
two separate CMEs interpreted as being responsible for two
adjacent MC events in the {\em Wind} data (events \#20-21 and
\#27-28).  Discussion of individual events is provided in an Appendix,
which should be consulted for more details of each particular
analysis.

     The $Q_s$ value listed in Table~2 is a quality flag indicating
our assessment of the level of evidence for an FR structure within each
CME, with $Q_s=1$ corresponding to no evidence for an FR shape, and
$Q_s=5$ corresponding to very strong evidence in favor of an FR shape.
Event \#5 is the only one at the negative $Q_s=1$ extreme.  This is a
faint event that is practically invisible in LASCO and COR1.  It appears
narrow and jet-like in the STEREO images at later times,
and it is simply too faint to provide any indication of what
its structure might be, FR or otherwise.

     There is also only one event that reaches the opposite $Q_s=5$
extreme, and that is event \#24, shown explicitly in Figure~3.  One of
the features we look for in searching for evidence of an FR shape is a
clear trailing edge for the FR, which is not always easy to discern.
For event \#24, there is a bright prominence eruption, most clearly
seen in the COR1 panels of Figure~3, which we believe can be
interpreted as nicely outlining the bottom of an FR that is viewed
face-on from the perspectives of STEREO-A and -B.  Even south of the
prominence, which fades quickly as the CME moves outwards, the HI1-B
image in Figure~3 shows a front that can be interpreted as the
trailing edge of the FR channel.  Another crucial characteristic that
we look for in assessing evidence for an FR shape is a clear dominant
axis to the structure, indicating the orientation of the FR.  The COR1
and COR2 images in Figure~3 make it very clear that this CME has a
wide angular extent in the north-south direction.  But from the
frontal perspective provided by the LASCO/C3 image in Figure~3, it is
clear that the CME is {\em not} very broad in the east-west direction.
This CME has a dominant north-south orientation, consistent with an FR
shape oriented roughly north-south.  The 2008~April~26 event studied
by \citet{bew09a} would be another $Q_s=5$ event, with a
circular appearance from STEREO-A's perspective suggesting an
FR structure viewed edge-on, consistent with the clearly dominant
east-west axis of the CME apparent from STEREO-B's frontal
perspective.

     Nearly half of our events are assigned the intermediate
$Q_s=3$ value, which corresponds to an event that we believe
we can model reasonably well with an FR shape, but which
we suspect could also be modeled equally well with some other
paradigm \citep[e.g., the traditional ``cone model'';][]{xpz02}.
There is of course significant subjectivity involved in our $Q_s$
index, but with only 7 of our 28 events having $Q_s=4$ or 5, we can
use it to state that even in the STEREO era only about a quarter of
CMEs in our estimation present an appearance in images that provides
clear positive evidence for an FR morphology, preferable to other
morphologies that might be assumed.  This could be used as an
argument against the FR paradigm being correct for all CMEs, but
the overall case for the FR paradigm outlined in Section~1 still seems
strong to us.  Therefore, we instead believe that our difficulty in
finding clear evidence for FR structures in all CMEs, even with
extensive STEREO and SOHO imaging constraints, is indicative of the
difficulties inherent in discerning magnetic structure from mass
distributions, which will not necessarily outline the underlying
magnetic structures very well.

\begin{figure}[t]
\plotfiddle{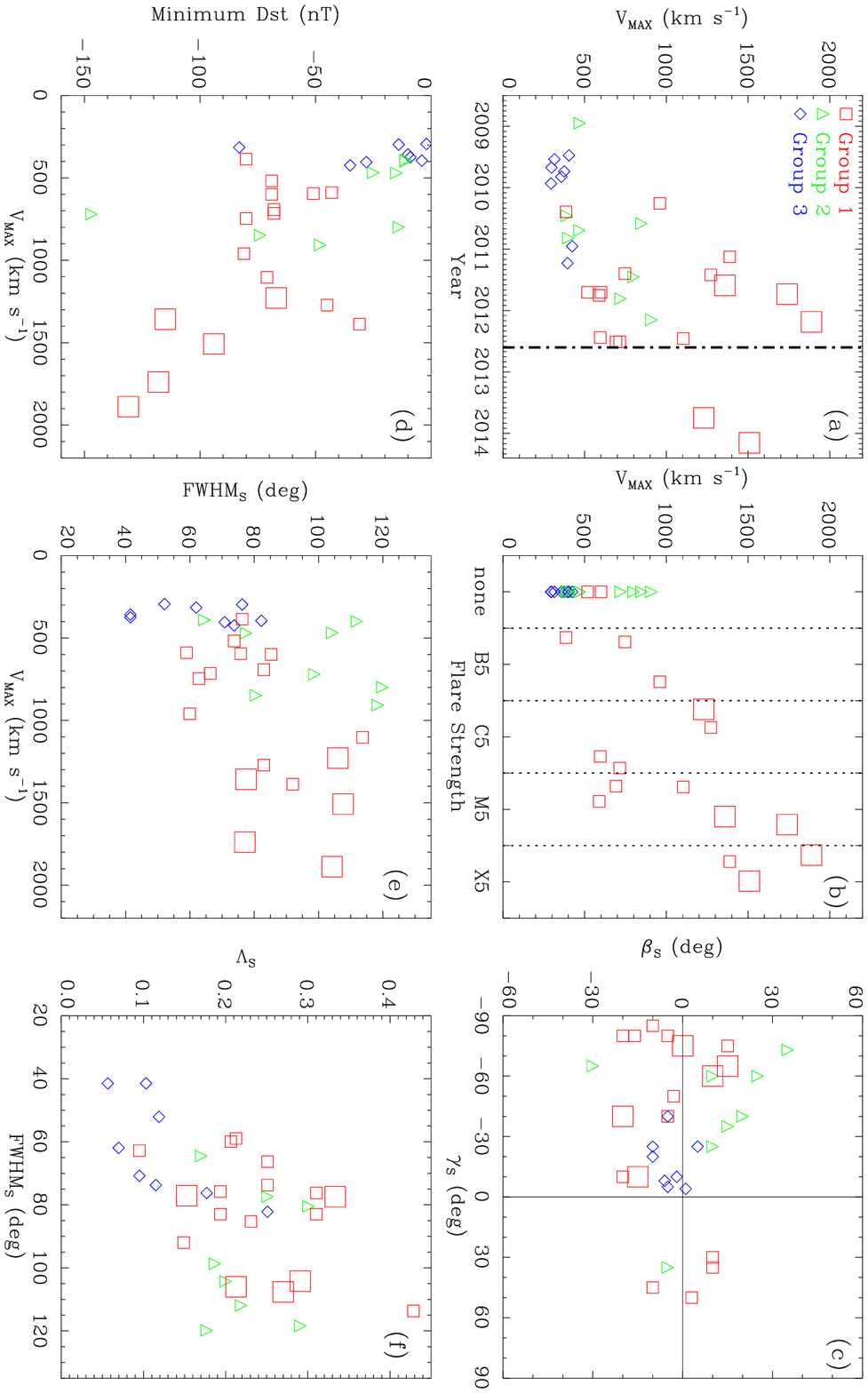}{4.4in}{90}{75}{75}{295}{-70}
\caption{Plots comparing various characteristics of the CMEs and
  their fitted FR parameters, with the CMEs divided into Group~1
  (flare-associated CMEs), Group~2 (filament eruption associated CMEs),
  and Group~3 (no surface activity).  The five large boxes are for the
  five added events from \citet{bew16b}.
  (a) Peak CME velocity plotted versus time.  The dot-dashed line
  indicates the end of the MC survey period.  (b) Peak velocity
  versus flare strength.  (c) Latitude of CME trajectory versus
  FR tilt angle.  (d) Minimum $D_{st}$ value versus peak velocity.
  (e) FR full-width-at-half-maximum versus peak velocity. (f) FR aspect
  ratio (the ``fatness'' of the FR) versus FR width.}
\end{figure}

\section{Characterizing Flux Rope Properties}

     Many of the measured CME and FR characteristics from Tables~1 and
2 are displayed in the various panels of Figure~7.  As in Figure~4(b) and
Figure~5(a), we find it instructive to divide our sample of events into
three groups according to associated surface activity, with group~1
being flare-associated events, group~2 being associated with filament
eruptions (but no flares), and group~3 associated with no clear surface
activity.  Also analogous to those figures, we include the five
additional events from \citet{bew16b}, which are geoeffective
events not perceived as MCs at Earth.

     Figure~7(a) illustrates how the character of our CMEs changes
significantly during the 2008--2012 period, indicating the shift from
solar minimum towards solar maximum.  The six 2009 events are all slow
group~3 CMEs.  In contrast, all but four of the events from 2011--2012
are fast group~1 events.  Figure~7(b) shows a clear correlation
between CME speed and flare strength, although there is still
significant scatter.

     The CME latitude and FR tilt angle are compared in Figure~7(c).
Curiously, 28 of the 33 events have negative tilt angles (i.e.,
$\gamma_s<0^{\circ}$).  In other words, the FRs tend to be tilted in a
northeast-southwest direction as opposed to northwest-southeast.  We
can think of no systematic error that would produce this, nor can we
think of any physical reason why there would be such a tendency.
Thus, if not simply a statistical fluke, the cause of this bias
remains a mystery.  In any case, one purpose of Figure~7(c) is to look
for a correlation between latitude and FR orientation.  Due to
differential rotation, filament channels in the northern hemisphere
tend to have negative tilt angles, and those in the southern
hemisphere tend to have positive tilts \citep{dmr94}.  Thus,
one might expect to see an overabundance of events in the upper left
and lower right quadrants of Figure~7(c).  This is definitely not the
case for the group~1 and group~3 events, but may be the case for
group~2 events, all but one of which are in the expected quadrants,
the exception being event \#13.  It is the filament eruption
associated group~2 events that would indeed most naturally be expected
to show this behavior.

     Figure~7(d) indicates some correlation between CME speed and
geoeffectiveness as measured by $D_{st}$.  The minimum $D_{st}$
(``disturbance storm time'') listed in Table~2
measures the depression of the horizontal component of the
geomagnetic field from a network of monitors near the equator, with
large negative values indicating a significant geomagnetic storm.
There is a great deal of scatter in Figure~7(d),
and no correlation at all would be apparent without
the addition of the 5 extra geoeffective events.  Event \#23 looks
like an outlier in this figure, with the most negative $D_{st}$ in the
figure despite a modest CME speed of only $V_{max}=720.3$ km~s$^{-1}$.

     Figure~7(e) plots FR width versus $V_{max}$.  The three groups
of CMEs are partly separated in this figure.  The slow group~3 FRs are
small, with $FWHM_s=63\pm 16$~deg.  For group~1, there is a clear
correlation between FR size and CME velocity.  For $V_{max}<1000$
km~s$^{-1}$, the mean width and standard deviation are $FWHM_s=71\pm
10$~deg; while for $V_{max}>1000$ km~s$^{-1}$, $FWHM_s=95\pm 15$~deg.
The group~2 FRs show a clear tendency to be larger than
group~1 and 2 FRs with similar velocities, with $FWHM_s=97\pm 21$~deg.
It is two group~2 events (\#19 and \#24; see Appendix) that are in
fact the largest FRs in our entire sample, suggesting that the largest
CME FRs are associated with filament eruptions from large filament
channels rather than X-ray bright active regions.  It is only the
shocks that make the fast $V_{max}>1000$ km~s$^{-1}$ CMEs appear to be
the largest of all CMEs.  Our analysis suggests that their FR ejecta
are not larger than the slower group~2 FRs.
Figure~7(f) shows the FR aspect ratio as a function of FR width.
There is a correlation, with the larger FRs being fatter, but it is
not a tight correlation.

     We find no correlation between the ellipticity of the FRs
($\eta_s$) and $V_{max}$, a correlation that one might expect if plowing
through a slower ambient solar wind yielded flattened FR channels.  In
particular, for fast events with $V_{max}>1000$ km~s$^{-1}$,
$\eta_s=1.45\pm 0.31$; little different from the value
for slow events with $V_{max}<1000$ km~s$^{-1}$, $\eta_s=1.42\pm 0.32$.
The very modest ellipticity that we infer for our events is not in
good agreement with past suggestions of substantial ``pancaking'' of
the FR during interplanetary travel, corresponding to $\eta_s\approx
4-6$ or more \citep{pr04,wbm04b,nps11}.
Our measurements would be in better agreement with
analyses implying or at least allowing for more modest flattening
\citep{mjo06,nps10}.

     Finally, we can compare our CME trajectory directions
in Table~2 ($\lambda_s$,$\beta_s$) with the locations of associated
surface activity, listed in Table~1 ($\lambda_p$,$\beta_p$)
The angular discrepancy between the two has a mean and standard
deviation of $19.3\pm 9.6$~deg.  Flares are much more compact than
filament eruptions, so group~1 events have a better defined source
location than group~2 events.  Considering group~1 alone, we find
a slightly lower angular discrepancy of $16.2\pm 8.1$~deg.  These
angular differences are generally interpreted as being due to
actual deflections in the CME's trajectory early in its journey from
the low corona \citep{ck13,ck15}.

\section{Comparing Imaging and In~Situ Perceptions of CME Structures}

     One of the main reasons to use in~situ {\em Wind} MC detections
to define our STEREO CME sample was to allow comparison of FR
characteristics inferred from imaging with MC characteristics
inferred from the {\em Wind} data.  Below we discuss three
different aspects of this comparison:  orientation, size, and
velocity/expansion.

\subsection{Orientation}

\begin{figure}[t]
\plotfiddle{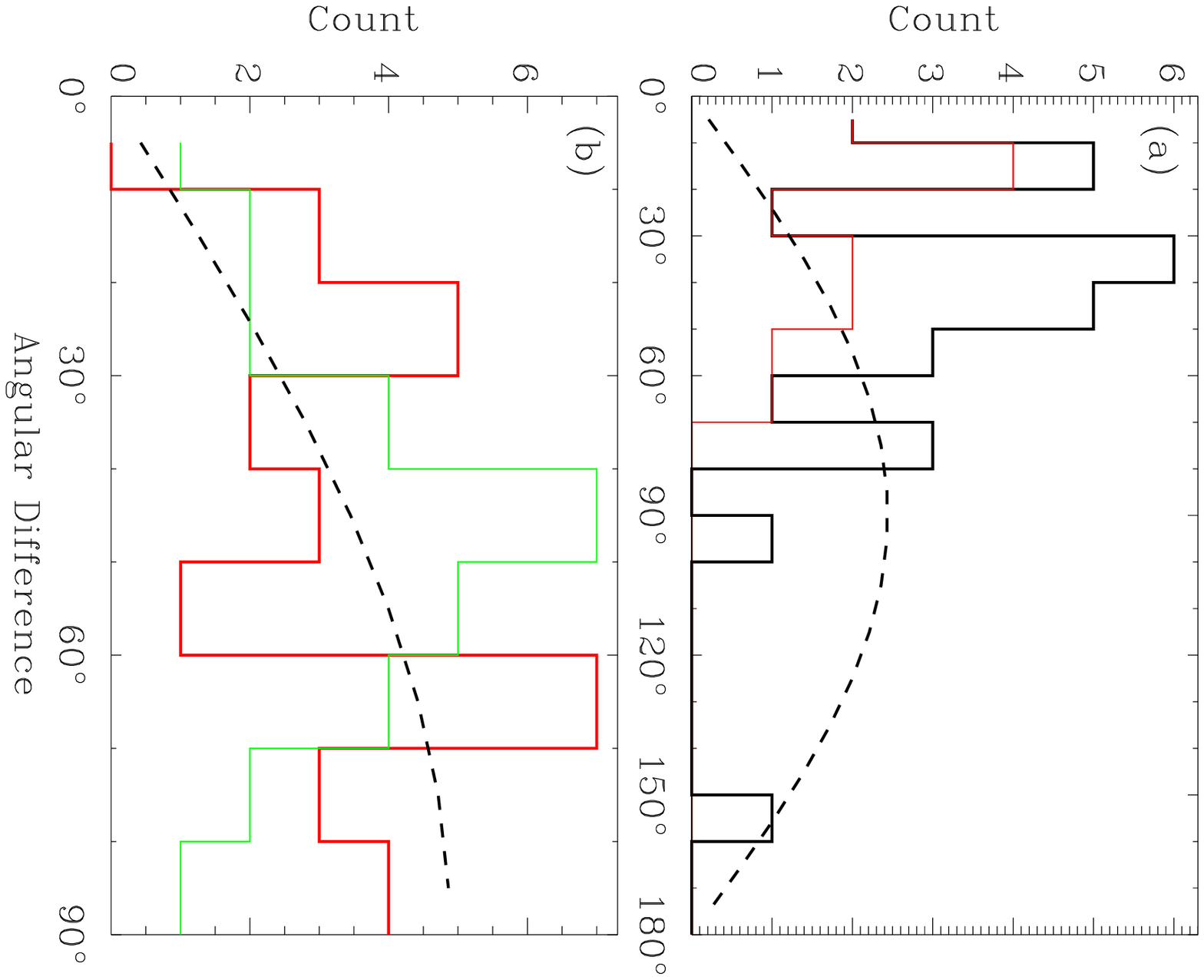}{3.6in}{90}{55}{55}{230}{-35}
\caption{(a) Distributions of angular discrepancies between the
  two independent measurements of MC axis direction listed in
  Table~1, ($\phi_l$,$\theta_l$) and ($\phi_n$,$\theta_n$), with
  the black line showing the full sample, and the red line
  excluding the poor $Q_l=3$ MCs.  The dashed line is the distribution
  expected for no correlation between the two measurements.
  (b) Distributions of angular discrepancies between the
  image-based FR axis direction from Table~2 ($\phi_s$,$\theta_s$)
  and the two MC axis direction measurements from Table~1,
  with the red line comparing with ($\phi_l$,$\theta_l$) and
  the green line with ($\phi_n$,$\theta_n$).  The dashed line
  is the distribution expected if there is no agreement, and the
  two computed distributions agree relatively well with this,
  suggesting inconsistency between the MC orientations and the
  FR orientations inferred from imaging.}
\end{figure}
     As discussed in Section~2, Table~1 provides two independent
assessments of MC orientation, both
assuming an infinite cylinder geometry with the cylinder axes pointing
towards ($\phi_l$,$\theta_l$) and ($\phi_n$,$\theta_n$) for the two
measurements, respectively.  We compute the angular discrepancy
between these two measurements, and Figure~8(a) shows the distribution
of these discrepancies.  Also shown is a line indicating what
the distribution would look like in the absence of any correlation,
showing the distribution of orientation differences expected for
two randomly oriented cylinders.  The agreement is certainly better
than random, but the average difference is still a rather large
$44^{\circ}$.  However, over half of our MC events have
been flagged as poor $Q_l=3$ MC events (see Table~1).
Figure~8(a) shows the distribution when the $Q_l=3$ events are
ignored.  This significantly improves the agreement, with the average
difference decreasing to $28^{\circ}$.

     However, the main goal here is to compare these MC orientations
with those of the FRs reconstructed from the images.  This is not
a trivial exercise, however, as the $\gamma_s$ tilt angle parameter
only indicates the FR axis orientation at the FR apex, and not
where {\em Wind} actually encounters the FR in our reconstruction.
Figure~6 shows Earth's path through the reconstructed FRs.  At each
point within the FR, we can find the shortest line connecting Earth
to the FR axis.  The plane normal to the FR axis containing this
line then defines the FR orientation at that point.  By computing
this orientation at all points along Earth's path through the FR,
and then averaging these orientations, we estimate an FR orientation
for Earth in the infinite cylinder approximation, appropriate for
comparison with the MC orientations.  These measurements
($\phi_s$,$\theta_s$) are listed in the last two columns of Table~2.
For the MC orientation, the magnetic field direction along the central
MC axis defines the direction quoted for the cylinder axis.  However,
images provide no such preferential direction, so both axis directions
are possible.  The direction quoted in Table~2 is the one with
$\phi_s<180^{\circ}$, but the other axis direction
($\phi_s+180^{\circ}$,$-\theta_s$) is also possible.

     Figure~8(b) compares the FR orientations with both MC
orientation measurements, and provides a line indicating what the
distribution should be in the absence of any correlation.  We have to
compare with both possible FR axis directions and take the minimum to
represent the discrepancy, meaning the worst possible agreement is
$90^{\circ}$ instead of $180^{\circ}$.  The two distributions are
similar to the non-correlation line, and the mean differences with
($\phi_l$,$\theta_l$) and ($\phi_n$,$\theta_n$) are $52^{\circ}$ and
$47^{\circ}$, respectively, only a little better than the $60^{\circ}$
value expected for no correlation.  We sought to improve agreement
by excluding $Q_l=3$ MCs and cases where the analysis is potentially
confused by multiple CMEs or consecutive MCs (see Appendix).  For
the remaining 9 events, the mean differences with
($\phi_l$,$\theta_l$) and ($\phi_n$,$\theta_n$) are $54^{\circ}$ and
$45^{\circ}$, respectively, showing no improvement over the full
sample.  Thus, we conclude that the image-based FR orientations are
simply not in good agreement with the MC orientations.

     Inferences of CME morphology from images and in~situ data both
carry substantial uncertainties that may be the cause of the poor
agreement seen in Figure~8(b).  Since images only show mass and
not magnetic field, the image-based analysis involves the
assumption that mass is loaded into the presumed FR in such a
way as to reveal its shape, but this may not always be the case.
The in~situ analyses are hampered by the fundamental difficulty of
inferring 3-D structure from a single, narrow, 1-D trace of plasma
properties through that structure.  There are also physical
assumptions involved that can be
overly simplistic (e.g.\ force-free, infinite cylinder, etc.).
It is possible to change these assumptions, in particular to
replace the cylindrical approximation with an elliptical
or toroidal geometry \citep{mah02,mv03,mv15,km07,mah12,nah13,qh14}.
It would be an
interesting exercise to see if such assumptions would yield MC
orientations in better agreement with the image-based inferences, but
such an analysis is beyond the scope of this paper.

     Although we do not in general claim to be certain at
this point whether the image-based or MC-based orientation
measurements are preferable, for the subset of events from 2009 we
would claim that the image-based orientations are more plausible.
The 2009 events are all slow events with no clear surface
activity associated with them.  They are all streamer
blowout type CMEs that develop under the existing quiescent
streamer structure, and therefore would be expected to have
their orientation determined by that of the overlying streamer belt.
All six of the 2009 CMEs are centered near the ecliptic plane,
and inspection of source surface synoptic charts computed from
magnetograms shows that the heliospheric current sheet was
relatively flat and near the ecliptic plane at the locations of
all six events.  Thus, the expectation is that all six FRs
should have a roughly east-west orientation,
meaning we should see $|\gamma_s|\approx 0^{\circ}$,
$|\theta_l|\approx 0^{\circ}$, and $|\theta_n|\approx 0^{\circ}$.
The actual measurements suggest averages and standard deviations of
$|\gamma_s|= 14.5\pm 13.7$~deg, $|\theta_l|=43.3\pm 15.7$~deg, and
$|\theta_n|=22.5\pm 22.8$~deg.  The image-based $\gamma_s$ values
best match the $0^{\circ}$ expectation, followed by $\theta_n$, and
lastly by $\theta_l$.  The fact that the ($\phi_n$,$\theta_n$)
measurements are noticeably more east-west than
($\phi_l$,$\theta_l$) could be evidence that the prescription for
relaxing the force-free assumption by \citet{tnc16}
leads to demonstrably more accurate MC analyses for these events.

\subsection{Size}

\begin{figure}[t]
\plotfiddle{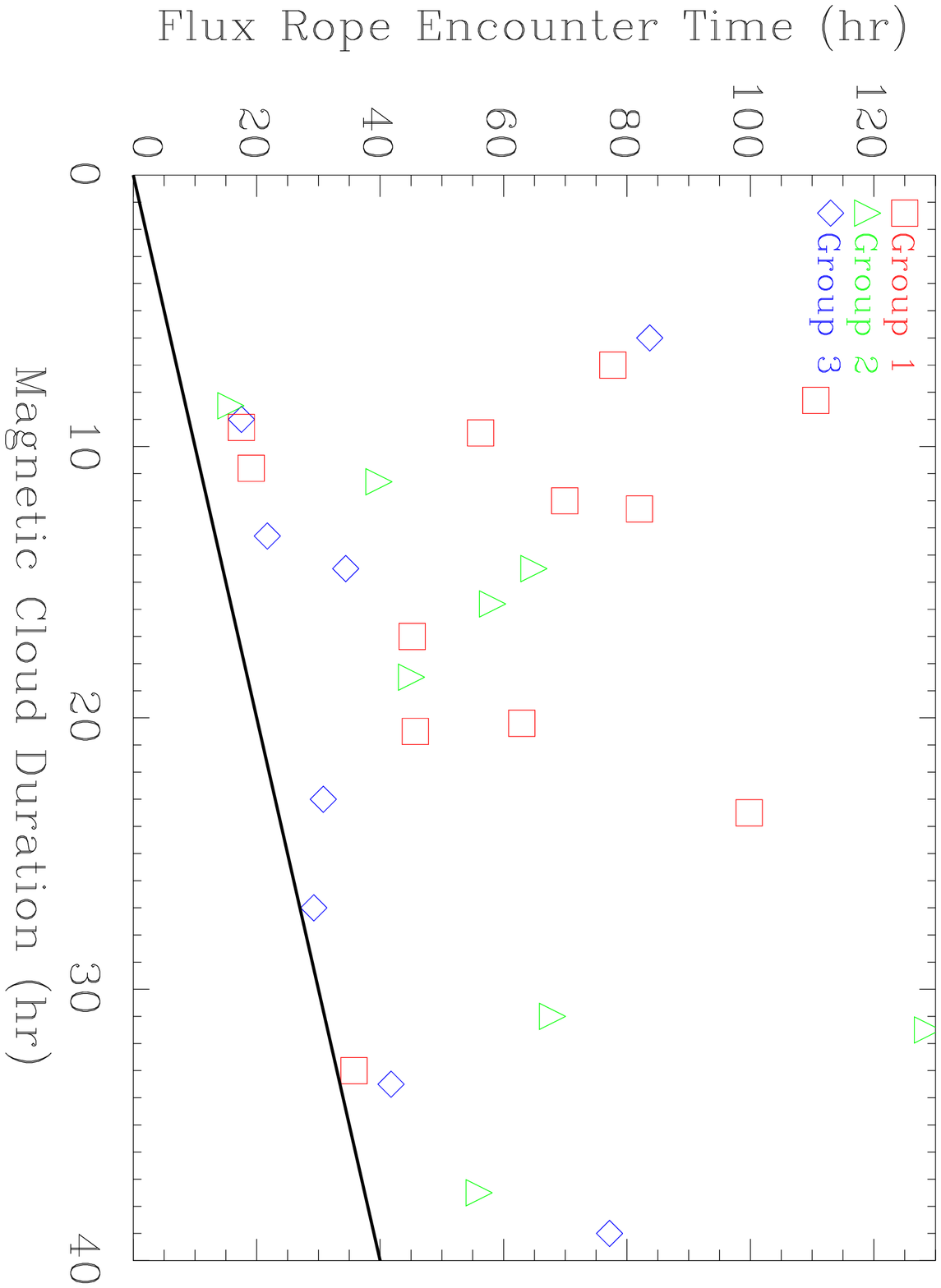}{3.5in}{90}{55}{55}{200}{-35}
\caption{Flux rope encounter time at {\em Wind} predicted by the
  CME reconstruction from imaging data, plotted versus magnetic cloud
  duration observed at {\em Wind}, with the solid line being the
  line of agreement.  Different symbols are used for Group~1
  (flare-associated CMEs), Group~2 (filament eruption associated CMEs),
  and Group~3 (no surface activity) events.}
\end{figure}
     Our image-based reconstruction of the FR morphology and kinematics
of the CME provides a prediction for how long {\em Wind} should
encounter the FR, and these FR encounter times can be compared with
the durations of the MCs observed by {\em Wind}.  This comparison is
made in Figure~1, which identifies the MC regions and also shows
for each event the density peaks predicted by the image-based
reconstruction, marking the predicted entry into and exit from
the FR.  (We are ignoring the shock density peaks for purposes of
this discussion.)  There are usually two separate peaks for the entry and
exit, except for grazing incidence cases that show only one peak
(e.g., event \#18).  In some cases, the exit density peak is not
visible in Figure~1 (e.g., event \#9) because it ends up beyond the
the 3.5~day period used in the figure.  These are typically cases where
the Earth and {\em Wind} travel into the leg of the FR, resulting in
a very late predicted exit time.

     It is very clear from cursory inspection of Figure~1 that the
predicted FR encounter times are generally much longer than the MC
durations.  In order to
make a more clear comparison, we compute FR encounter times from the
density peaks.  For purposes of this calculation, we define the FR
encounter time as follows.  For each event we define a somewhat
arbitrary density threshold of 15\% of the maximum density, and we
define the entry time as being when this threshold is first reached,
and the exit time as when this density is last encountered.

     The resulting FR encounter times are compared with the MC
durations in Figure~9, where we once again divide the events into the
three groups used in previous figures.  It should be emphasized that
some of these FR encounter times are surely unreasonably long.  In
the self-similar expansion approximation used in the FR reconstruction,
the velocity at a point within the FR is proportional to its distance
from the Sun.  Thus, if {\em Wind} is predicted to exit from the leg
of an FR at a third of the distance to the FR's leading edge, the
velocity at that point will only be a third of the leading edge
velocity, which is likely to be unreasonably slow in most cases,
leading to an implausibly long FR encounter time.
In reality, the ambient solar wind will be impinging on the trailing
parts of the FR, presumably accelerating them to higher velocities
than our reconstructions would suggest.

     With this caveat in mind, Figure~9 illustrates the degree to
which the FR encounter times are consistently longer than the MCs.
The average MC duration is 18.5~hr, compared with 54.8~hr for the
average FR encounter time.  The FRs inferred from
imaging are simply much fatter than would be expected from the
MC durations.  Not surprisingly, the disagreement is less acute for
the group~3 events, which are the narrowest FRs, having the lowest
$\Lambda_s$ values.  The substantial FR/MC size discrepancy
suggests that there is not a clear one-to-one correspondence between
the FR channels inferred from imaging and those seen as MCs by
in~situ instruments.  Perhaps the true magnetic flux ropes are the
narrow structures seen as MCs, but these FRs exist within
larger ICME channels outlined by messier, more irregular magnetic
barriers, representing the channels perceived as FR shapes in images.
This could also explain the orientation differences between MCs and
image-based FRs, as the MC orientation would not necessarily be
identical to that of the larger ICME channel.

\subsection{Velocity and Expansion}

\begin{figure}[t]
\plotfiddle{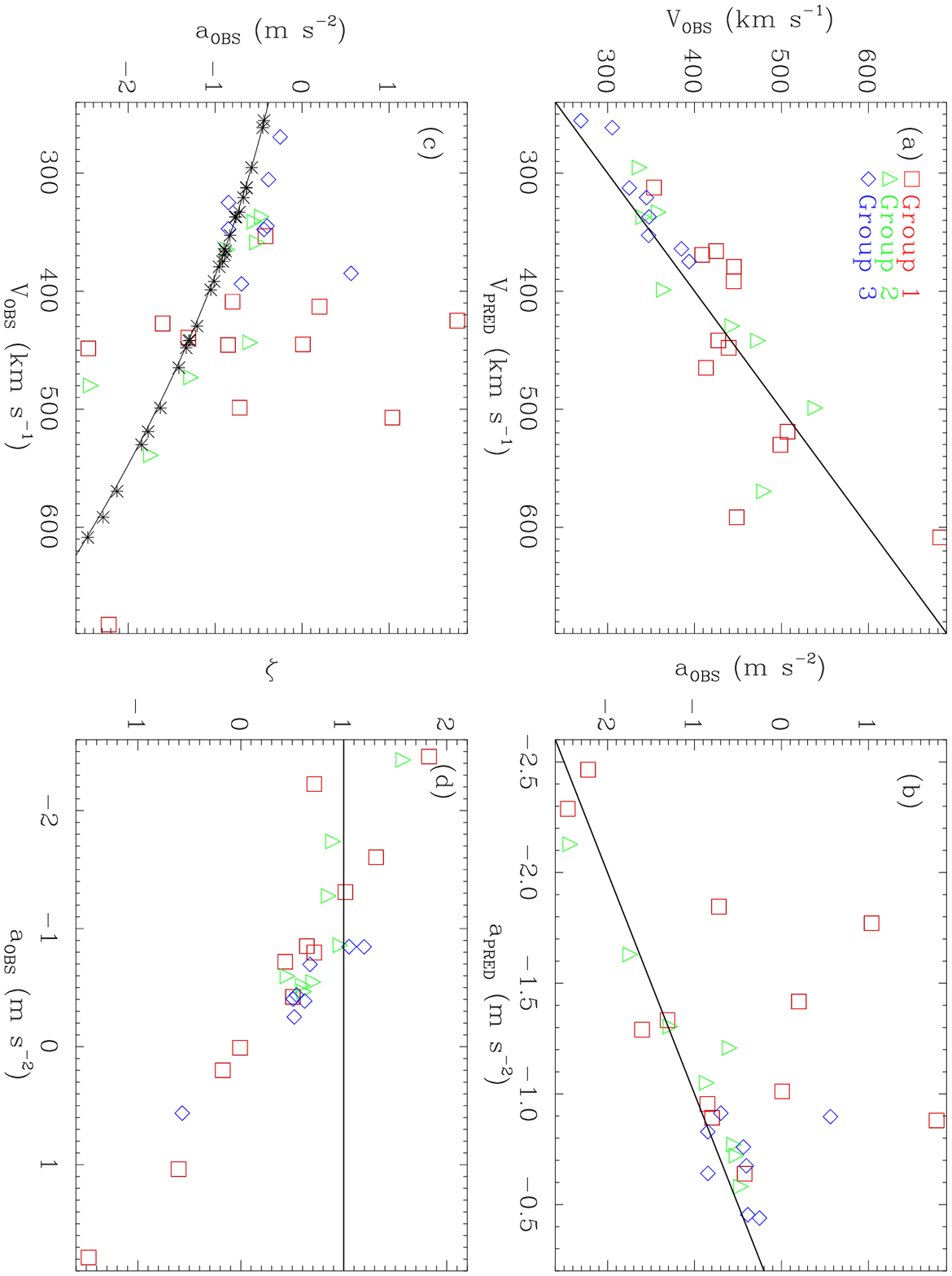}{4.6in}{90}{70}{70}{275}{-45}
\caption{(a) Observed average velocity within the MC compared
  with that predicted by the image-based CME reconstruction, with the
  solid line being the line of agreement.  Different symbols are used
  for Group~1 (flare-associated CMEs), Group~2 (filament eruption
  associated CMEs), and Group~3 (no surface activity) events.
  (b) Observed average acceleration within the MC compared
  with that predicted by the image-based CME reconstruction, with the
  solid line being the line of agreement.  (c) Acceleration
  within the MC plotted versus velocity.  The asterisks show the
  predicted instead of observed values, which follow the expected
  $a_{obs}=-V_{obs}^2/d$ relation for self-similar expansion.  (d) The
  dimensionless expansion factor $\zeta$ versus acceleration within
  the MC.}
\end{figure}
     Figure~1 compares velocities observed by {\em Wind} with
velocities predicted by the image-based reconstructions.  But
comparing the velocities requires some care.  This is especially true
with regards to shock and shock sheath regions.  The images will be
indicating the shock propagation speed while the speeds measured by
{\em Wind} will be the velocities of material moving through the
shock, and these are not necessarily the same things (see discussion
of event \#24 in the Appendix).  The MC regions identified for our
events should be CME ejecta, however, where the predicted velocities
should unambiguously be in agreement with the observed velocities.
Therefore, we here compare the predicted and observed velocities
within the MC regions.

     We compute mean observed and predicted velocities
within the MC bounds ($V_{obs}$ and $V_{pred}$, respectively), and
compare them in Figure~10(a).  The observed velocities are in
reasonably good agreement with those predicted by the image-based
reconstruction, with an average discrepancy of only $\pm 37$ km~s$^{-1}$.
This provides encouraging support for the accuracy of the kinematic
aspects of the 3-D reconstructions.  However, there are some curious
systematic discrepancies.  In particular, the reconstructions clearly
tend to underpredict velocities for the slower CMEs, perhaps due to
perturbations of the MC resulting in incursions of higher velocity
material into the MC region.

     The predicted velocities in Figure~1 all have negative
slopes.  This is a consequence of the self-similar expansion
assumption used in the FR reconstruction process.  With this
assumption, velocity within the structure is proportional to distance
from the Sun.  Thus, trailing parts of the CME are slower, so velocity
decreases with time.  Inspection of the MC regions in Figure~1 shows
that most do in fact show observed velocity decreases, qualitatively
consistent with expectations.  We perform linear fits to the observed
and predicted velocities within the MC bounds, with the slopes of the
resulting lines indicating accelerations seen at {\em Wind} ($a_{obs}$
and $a_{pred}$, respectively).  These accelerations are compared in
Figure~10(b).  There is reasonable agreement for most events, with
$a_{obs}$ falling within $\pm 0.35$ m~s$^{-2}$ of the prediction for
75\% of the events, but there are a number of outliers, particularly
five events that actually have positive $a_{obs}$, suggesting a contracting
MC instead of an expanding one.

     The nature of the MC expansion can be explored further
by combining the $V_{obs}$ and $a_{obs}$ measurements.  These two
quantities are compared in Figure~10(c).  For self-similar expansion,
velocity and acceleration are related by $a_{obs}=-V_{obs}^2/d$ (see
below), where $d$ is the distance from the Sun, and Figure~10(c)
demonstrates that the measured $a_{pred}$ and $V_{pred}$ points
naturally fit this relation.  Most of the observed data points fall
above this line, suggesting the observed MCs are not generally
expanding as fast as the self-similar expansion approximation would
predict.  This is better explored using the dimensionless expansion
parameter
\begin{equation}
\zeta=-\frac{a_{obs} d}{V_{obs}^2},
\end{equation}
defined by \citet{pd08} and \citet{pd09}.
For self-similar expansion, $\zeta=1$.  Values of $\zeta>1$
indicate expansion faster than self-similar, $\zeta=0-1$ indicates
expansion slower than self-similar, and $\zeta<0$ indicates
contraction.  Note that $\zeta$ is essentially identical to the $S_r$
self-similarity factor derived by \citet{bew16a} in the context
of image analysis.

     Figure~10(d) plots $\zeta$ versus $a_{obs}$.  Exactly
75\% of the points fall in the range $\zeta=0.5-1.5$, which could be
described as crudely self-similar.  However, most of these points lie
below $\zeta=1$.  This is consistent with the measurements of
\citet{pd08}.  \citet{amg10,amg12} find that
MCs with relatively linear velocity trends, which they refer to as
``non-perturbed'' MCs, tend to have $\zeta\approx 1$.  It is clear
that many of our MCs would be considered ``perturbed'' in this
regard, potentially accounting for the $\zeta<1$ tendency.

\section{Predicting 1 AU Arrival Times}

\begin{figure}[t]
\plotfiddle{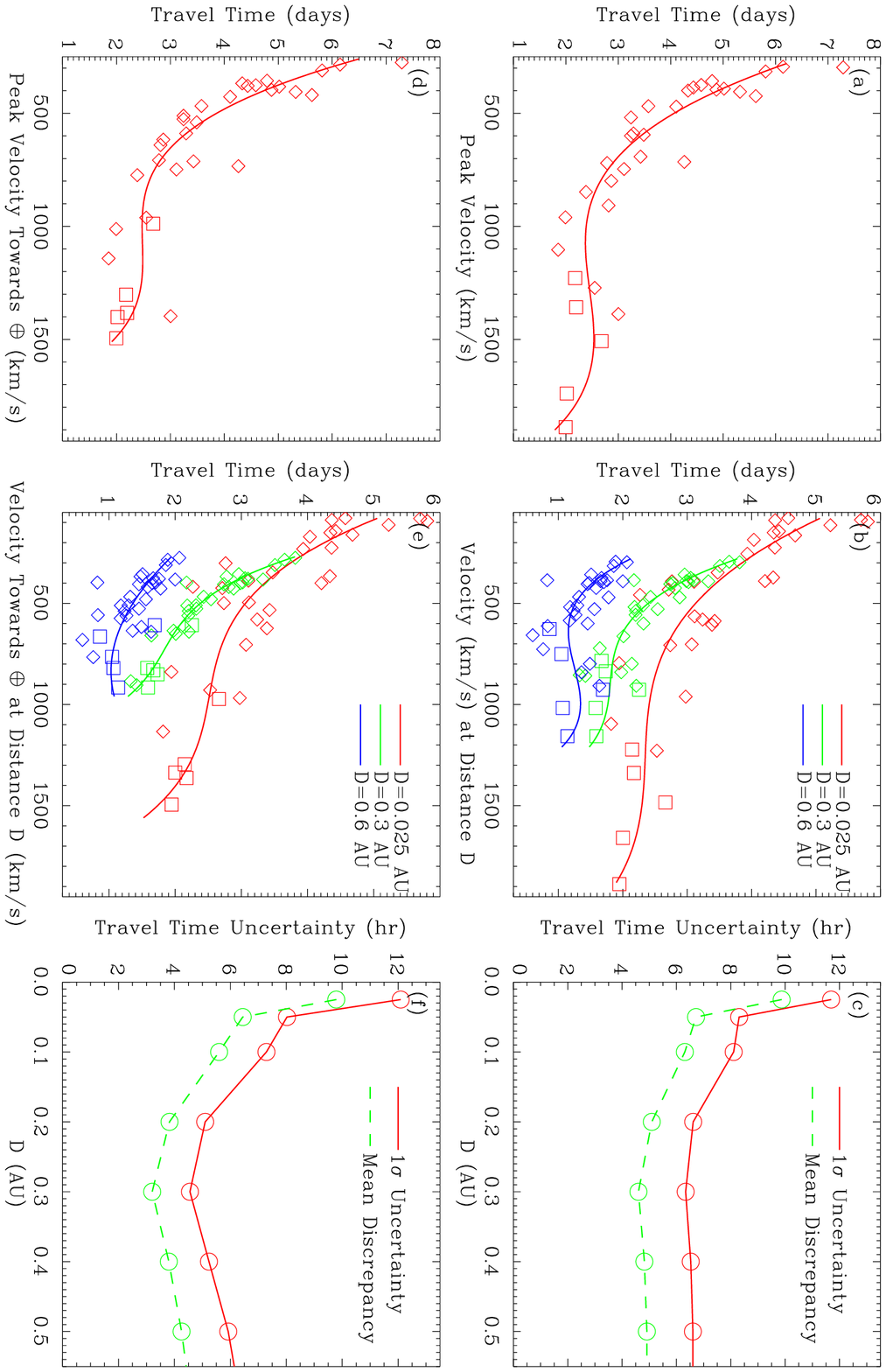}{4.4in}{90}{75}{75}{290}{-70}
\caption{(a) Travel time from the Sun to Earth as a function of peak
  CME velocity at its leading edge.  Diamonds are the MC sample of
  events from Table~1, and boxes are the five included events from
  \citet{bew16b}.  A third-order polynomial is fitted to the
  data points.  (b) Travel time from distance D to Earth as a
  function of CME leading edge velocity measured at distance D, shown
  for 3 values of D.  Third-order polynomials are fitted to the
  curves, with the fit coefficients listed in Table~3.
  (c) Uncertainties in travel time estimates to Earth as a function
  of distance where a CME's leading edge velocity is measured,
  computed from the scatter about curves like those in
  (b).  Panels (d)-(f) are analogous to panels (a)-(c), but instead of
  leading edge velocity, the velocity used is the velocity of the
  part of the CME moving towards Earth.}
\end{figure}
     The kinematic profiles in Figure~4(b) represent a significant
improvement in our understanding of how CMEs propagate from the
Sun to 1~AU, being constrained both by the observed Earth arrival
time, and extensive STEREO monitoring continuously from the Sun
to 1~AU.  Since we know the ICME arrival times at {\em Wind} for all
33 of these events, these measurements can be used to explore how CME
velocity at a given distance from the Sun relates to remaining travel
time to Earth.   This CME arrival time forecasting is useful
for operational space weather forecasting purposes.  Thus, we here
explore the implications of our measurements for Earth arrival time
predictions.

\begin{table}[t]
\scriptsize
\begin{center}
\caption{Relating CME Velocity to 1~AU Travel Time}
\begin{tabular}{ccccccccc} \hline \hline
 & & & \multicolumn{4}{c}{Coefficients$^c$} & & \\
\cline{4-7}
 $D$$^a$ & $V_-$$^b$  &  $V_+$$^b$ & $a_0$ &   $a_1$    &   $a_2$    &
     $a_3$   & $\overline{\delta t_{AU}}$$^d$ & $\sigma_{t}$$^e$ \\
(AU) & (km s$^{-1}$) & (km s$^{-1}$) &      & (10$^{-3}$) & (10$^{-6}$) &
  (10$^{-9}$) &       (hr)       &       (hr)       \\
\hline
\multicolumn{9}{l}{\underline{Analysis with Leading Edge Velocities}} \\
0.025&  78.7 & 1887.7 & 5.63 & -7.55 & 5.88 & -1.55 & 9.9 &11.7 \\
0.05 & 141.5 & 1861.1 & 6.00 & -7.64 & 5.05 & -1.13 & 6.7 & 8.3 \\
0.10 & 203.9 & 1593.8 & 6.51 & -9.53 & 6.69 & -1.55 & 6.3 & 8.1 \\
0.20 & 291.0 & 1156.6 & 7.90 & -17.4 & 17.1 & -5.67 & 5.1 & 6.6 \\
0.30 & 294.2 & 1156.6 & 7.59 & -18.8 & 20.6 & -7.67 & 4.6 & 6.3 \\
0.40 & 294.2 & 1156.6 & 7.05 & -19.1 & 22.2 & -8.54 & 4.8 & 6.5 \\
0.50 & 294.2 & 1156.6 & 6.29 & -18.4 & 22.7 & -9.07 & 4.9 & 6.6 \\
0.60 & 294.2 & 1156.6 & 5.18 & -16.0 & 20.6 & -8.38 & 4.9 & 6.6 \\
\multicolumn{9}{l}{\underline{Analysis with Velocities Towards Earth}} \\
0.025&  79.0 & 1494.7 & 5.71 & -8.87 & 8.68 & -3.02 & 9.8 &12.1 \\
0.05 & 136.8 & 1473.7 & 6.00 & -7.78 & 4.99 & -1.05 & 6.4 & 8.0 \\
0.10 & 197.2 & 1262.0 & 6.38 & -8.67 & 4.59 & -0.38 & 5.6 & 7.3 \\
0.20 & 275.0 & 1065.9 & 7.66 & -16.3 & 15.2 & -4.86 & 3.8 & 5.1 \\
0.30 & 275.0 &  915.8 & 7.80 & -21.1 & 26.5 & -12.1 & 3.2 & 4.6 \\
0.40 & 275.0 &  915.8 & 5.67 & -12.4 & 12.6 & -4.74 & 3.8 & 5.2 \\
0.50 & 275.0 &  915.8 & 4.37 & -8.60 & 7.58 & -2.18 & 4.3 & 5.9 \\
0.60 & 275.0 &  915.8 & 2.96 & -4.37 & 2.18 &  0.35 & 4.6 & 6.4 \\
\hline
\end{tabular}
\end{center}
\tablecomments{$^a$Distance where velocity measurement made.
  $^b$($V_-$,$V_+$) is the allowable range of velocities.
  $^c$Coefficients of the third order polynomial fit to compute the 1~AU
  travel time (see equation (6)).  $^d$Mean discrepancy
  with the Earth arrival time prediction.  $^e$Standard deviation of
  discrepancies with the Earth arrival time prediction.}
\normalsize
\end{table}
     Figure~11(a) shows how the total Sun-to-Earth travel time
correlates with peak CME speed.  The slowest events in our sample,
which reach only $V_{max}\approx 300$ km~s$^{-1}$, take $6-7$ days to
make it to Earth.  The fastest events with $V_{max}>1000$ km~s$^{-1}$
take only about $2-3$ days.  However, for
forecasting purposes, it is not {\em peak} CME speed that one wants
to relate to 1~AU travel time, but CME speed at a particular height.
We can use the kinematic models in Figure~4(b) to tell us the
leading edge speed of all of the CMEs at any distance $D$, and knowing
the ICME arrival times at Earth, we can compare these speeds with the
remaining travel time from distance $D$ to Earth.  This is done in
Figure~11(b) for $D=0.025$, 0.3, and 0.6~AU.  We fit these curves
with third order polynomials, such that the 1~AU travel time (in
days) is
\begin{equation}
t_{AU}=\sum^3_{i=0} a_i V^i,
\end{equation}
where $V$ is the velocity at distance $D$ in km~s$^{-1}$.  The
$a_i$ coefficients are listed in Table~3 for a range of $D$
values between 0.025 and 0.6~AU, allowing travel times to Earth
to be predicted from a measured CME velocity anywhere in this
distance range.  This prescription will only apply within the
velocity range covered by our sample of events, which is
given by the ($V_-$,$V_+$) values listed in Table~3.

     The scatter about the polynomial fits in Figure~11(b) can
be used to estimate uncertainties in the $t_{AU}$ predictions.  The
most natural estimate is the standard deviation of the discrepancies
from the fit, $\sigma_t$, which is listed in the last column of
Table~3.  But in the literature, sometimes the mean discrepancy,
$\overline{\delta t_{AU}}$, is provided instead, so we also list this
value in the table.  Figure~11(c) plots both $\sigma_t$ and
$\overline{\delta t_{AU}}$ versus $D$.  Close to the Sun at
$D=0.025$~AU the travel time uncertainty is $\sigma_t=11.7$~hr.  This
value quickly drops as $D$ increases, reaching a minimum at about
$D=0.3$~AU, where $\sigma_t=6.3$~hr (and $\overline{\delta
t_{AU}}=4.6$~hr).  The reason for the minimum at $D=0.3$~AU is that
by this distance most CMEs have finally reached their final, terminal
speeds, with velocities that do not change much after this (see
Figures~4(b) and 5(b)).  The uncertainties do not decrease any further
with increasing $D$.

     The velocities shown in Figure~4(b) and used in the
analysis depicted in Figures~11(a-c) are leading edge velocities.  One
potentially significant contributor to the uncertainties $\sigma_t$
and $\overline{\delta t_{AU}}$ is that Earth is not always hit near
the leading edge of a CME.  In many cases Earth is hit by a part of
the CME front well behind the leading edge, and therefore traveling
slower.  Because our analysis involves a full morphological
reconstruction of the CMEs, we know what the actual CME speeds towards
Earth are, and we can make a version of Figure~4(b) with those
velocities instead of the leading edge velocities.  Figure~11(d-f)
repeats the analysis of Figure~11(a-c), but using these Earth-directed
CME speeds, in order to see if this significantly reduces
uncertainties.  The results, which are also listed in Table~3, are
unexpected.  There is significant decrease in $\sigma_t$ and
$\overline{\delta t_{AU}}$ only at $D=0.2-0.4$~AU.  At $D=0.3$~AU, the
uncertainties dip to an impressively low $\sigma_t=4.6$~hr (and
$\overline{\delta t_{AU}}=3.2$~hr), but they then actually increase
for $D>0.3$~AU, which is counterintuitive.

     In order to explain this, we note that there are two
primary sources of uncertainty: (A) the uncertainty in the velocity
measurements providing the basis for this calculation (i.e.,
Figure~4(b)); and (B) the intrinsic variation in 1~AU travel time for a
set of CMEs at a given velocity at a given distance D, which may have
different 1~AU travel times despite the same starting conditions, due
to different solar wind conditions between distance D and Earth.  The
increase in uncertainty for $D>0.3$~AU seen in Figure~11(f) can really
only be explained by option (A).  It is quite possible that our
velocity measurements are more uncertain for $D>0.3$~AU.  This may be
because measurements from the HI2 field of view are simply
intrinsically more uncertain, or it could be an artifact of our
particular method of velocity measurement, assuming a few phases of
constant velocity or acceleration, and always assuming a final,
relatively lengthy phase of constant velocity (see Section~3.1).

     In practice, accurately estimating a Sun-Earth line CME velocity
in a real time forecasting situation will not be easy, unless one
actually has a spacecraft operating much closer to the Sun on the
Sun-Earth line.  For now, the limited improvement in $\sigma_t$ and
$\overline{\delta t_{AU}}$ is probably not sufficient to be worth the
extra effort.  We suspect that using the leading edge velocities is more
practical.  Though even here we should emphasize that an effort must
be made to estimate the true 3-D leading edge velocity of the CME, and
not just a projected velocity in an image plane, which would not be
appropriate to use in equation (6).

     In principle, the arrival time uncertainties of our
proposed forecasting prescription can be compared with other
published assessments \citep{ng00,ng01,ekjk12,gm13,cm14,bv14,mlm15,tr16},
but in practice this is not trivial, as different analyses assume
different constraints available at different distances from the Sun,
and use samples of CMEs with different distributions of fast and slow
events.  Some quoted 1~AU travel time uncertainties are for
predictions for the full travel time from near the Sun to 1~AU, and
some are for predictions after the CME is already within a day or so
of reaching Earth.  The last two columns of Table~3 by themselves
represent many different quantifications for 1~AU travel time
forecasting accuracy.  We believe the uncertainties quoted for our
prescription in Table~3 are competitively low compared to ones cited
in the papers cited above.  But it should be emphasized that in an
actual real time forecast, these uncertainties would have to be added
in quadrature with the errors introduced by the uncertainty in the
velocity measured for that particular event, and it is that
uncertainty that may end up dominating the accuracy of the 1~AU
travel time estimate.

\section{Summary}

     We have conducted a survey of MC CMEs observed by {\em Wind}
from 2008 to mid-2012 that can be tracked from the Sun to near 1~AU
by STEREO, performing a 3-D reconstruction of each event's morphology
and kinematic behavior based on STEREO and SOHO/LASCO imagery.  Our
principle results are summarized as follows:
\begin{enumerate}
\item Out of 48 {\em Wind} MCs, 31 are definitively connected with
  CMEs trackable to near 1~AU by STEREO.  Most of the remaining
  17 unaffiliated MCs are probably CMEs as well, but they may be
  too faint or confused with other CME/CIR fronts in the images
  for a definitive connection.
\item Our 3-D CME reconstruction approach assumes an FR morphology.
  We find the FR paradigm to be reasonably successful at reproducing
  CME appearance in the images, but in our judgment only about a
  quarter of the events provide sufficient evidence for an FR shape to
  suggest that the FR paradigm is clearly preferable to other
  paradigms for CME morphology (e.g., the cone model).
\item The FR orientations from our image-based reconstructions are
  not in good agreement with MC orientations inferred from {\em Wind}
  data, and the FRs are usually too large to be consistent with MC
  durations.
\item Focusing only on six streamer blowout CMEs from 2009, which
  should have roughly east-west FR orientations, the image-based
  reconstructions generally recover the expected orientations.  The MC
  analyses of the {\em Wind} data are not as successful in this
  regard, but there seems to be significant improvement when the
  traditional force-free assumption is relaxed.
\item The image-based reconstructions reproduce the observed mean
  velocities within MC regions to within $\pm 37$ km~s$^{-1}$ on
  average.  The reconstructions assume self-similar expansion.
  Exactly 75\% of the MCs exhibit accelerations crudely consistent
  with self-similar expansion ($\zeta=0.5-1.5$), though there is
  a clear tendency for slower expansion rates ($\zeta < 1$),
  probably due to perturbations of the MCs.
\item Our reconstructions provide full Sun-to-Earth kinematic
  measurements.  After supplementing our main sample with five
  geoeffective events from \citet{bew16b}, we find that
  14 of 16 CMEs with $V_{max}>600$ km~s$^{-1}$ have measurable
  deceleration in the interplanetary medium after reaching their
  peak velocity.
\item We use the kinematic models to provide simple prescriptions
  for predicting CME arrival times at Earth.  These prescriptions
  are provided for a range of distances from the Sun where CME
  velocity measurements might be made ($D=0.025-0.6$~AU).
  The 1$\sigma$ uncertainties in these predictions range from
  $\sigma_t=4.6-11.7$~hr (or mean discrepancies of
  $\overline{\delta t_{AU}}=3.2-9.9$~hr), depending on the distance
  $D$ of the velocity measurement and on whether that velocity is
  a leading edge velocity or a velocity along the Sun-Earth line.
\item We find it instructive to group our events based on
  associated surface activity:  group~1 (flare-associated CMEs),
  group~2 (filament eruption associated CMEs), and group~3 (no
  surface activity).  These groups exhibit noticeably different
  kinematic behavior, with clear differences in peak velocities
  ($V_{max}$) and heights where $V_{max}$ is first reached.
  For group~3, $V_{max}=294.2-423.8$ km~s$^{-1}$ and
  $R=29.4\pm 14.8$~R$_{\odot}$; while for group~1, $V_{max}>500$ km~s$^{-1}$
  for all but one event, and $R=3.2\pm 2.2$~R$_{\odot}$.  The
  group~2 events are intermediate, with $V_{max}=391.0-907.6$ km~s$^{-1}$
  and $R=13.9\pm 3.8$~R$_{\odot}$.
\item The three groups are partially separated in parameter space
  when FR width is plotted versus $V_{max}$, with group~3 events
  being slow and relatively narrow, group~1 events showing a
  correlation of width with $V_{max}$, and group~2 events being
  generally broader than the other events at similar speeds.
  Our two broadest modeled FRs are both group~2 events.
\end{enumerate}

\acknowledgments

Financial support was provided by the Chief of Naval Research,
and by NASA awards NNH10AN83I and NNH14AX61I to the Naval Research
Laboratory.  TNC is partially supported by the National Science
Foundation under AGS-1433086 and the NASA/LWS program. The
STEREO/SECCHI data are produced by a consortium of NRL (US), LMSAL
(US), NASA/GSFC (US), RAL (UK), UBHAM (UK), MPS (Germany), CSL
(Belgium), IOTA (France), and IAS (France).
We acknowledge use of NASA/GSFC's Space Physics Data Facility's
OMNIWeb (or CDAWeb or ftp) service, and OMNI data.  The $D_{st}$
values used in this paper are from Kyoto's World Data Center for
Geomagnetism (http://wdc.kugi.kyoto-u.ac.jp/dstdir/index.html).

\appendix
\section{Notes on Individual Events}

     We here provide a brief discussion of the individual
events, particularly important for cases in which multiple
CMEs make interpretation of Figure~1 and Figure~6 difficult.
Movies of each event showing the STEREO and SOHO/LASCO images of
these CMEs, compared with synthetic images from the CME
reconstructions described above, are provided in the online
version of this article.
\begin{description}
\item[Event \#1] This is a well-studied event that was the first
  clear case of an Earth-directed CME continuously trackable all
  the way from the Sun to 1~AU \citep{cjd09,yl10,ced11}.
  Our analysis has the FR hitting Earth
  near our estimated ICME arrival time, which is about 18 hours
  earlier than the MC start time (see Figure~1).
\item[Event \#2] This is a CME that changes appearance significantly
  as it expands away from the Sun, meaning the self-similar expansion
  approximation is unusually poor in this case.  We have already
  published a description of a more complex, time-dependent analysis
  of this event \citep{bew12a}, to which we refer the reader for
  details.
\item[Event \#3] This is a faint, narrow CME from the perspective
  of both STEREO-A and -B, which is undetectable in the SOHO/LASCO
  images.  The in~situ ICME signature, shown in more detail by
  \citet{ekjk14}, is probably the weakest of all
  our events, with only a very small density and field increase.  A
  CIR arrives a day after our estimated ICME arrival time, producing
  far more dramatic density, velocity, and field increases.
\item[Event \#4] This is a particularly instructive case for
  illustrating the value of multiple viewpoints for correct
  interpretion.  As shown in Figure~6, our analysis reveals two
  separate CMEs, the first (CME1) beginning at about 15:00 on
  2009~September~2, directed west of Earth, which hits STEREO-A on
  September~8.  It is the second (CME2), which starts about 10 hours
  later, that hits Earth, though the center of the CME is well to the
  east of Earth.  From SOHO/LASCO's perspective, CME1 is readily
  apparent, but CME2 is not clearly discernible as a separate eruption.
  From STEREO-A's perspective the situation is reversed --- CME1 is
  nearly invisible, and what evidence there is of
  it is not easy to discern as being distinct from CME2, which is
  relatively bright.  It is only STEREO-B that provides a perspective
  that makes it clear that there are two distinct slow streamer
  blowout CMEs.  This may be a case where the first streamer blowout
  CME triggered an eruption from an adjacent part of the streamer
  belt, resulting in CME2, which ultimately yielded the MC at Earth.
  This event is also notable for having the longest 1~AU travel time
  in our sample, taking over 7 days to get from the Sun to Earth.
\item[Event \#5] We have already mentioned in Section~3.2 how this
  particularly faint, jet-like event is the only $Q_s=1$ CME in our
  sample, providing no evidence of an FR morphology, although
  the reconstructed FR nevertheless ends up matching the MC time
  range well in Figure~1.
\item[Event \#6] This is another faint, streamer blowout CME from
  2009, with only very weak signatures in SOHO/LASCO.
  In COR1, the CME reaches the edge of the COR1-B field of
  view well before it reaches the edge of COR1-A, but this changes
  by the time the CME exits COR2, suggesting the CME may have started
  out with a more westerly trajectory, quickly deflecting eastwards
  in the COR2 field of view.  We cannot reproduce this in our
  reconstruction without abandoning the self-similar expansion
  approximation.
\item[Event \#7] This CME seems to be part of a general outflow from
  part of the streamer belt, which makes it very difficult to decide
  what part of this outflow represents the leading edge of the CME.
  The Earth arrival time constraint provided by the {\em Wind}
  data is crucial for deciding what front to follow as
  the leading edge.  Unlike many of the other 2009 streamer blowout
  CMEs, there is actually a nice SOHO/LASCO partial halo for this
  event, which is very helpful for constraining the FR orientation.
\item[Event \#8] This is a very well studied event, notable for being
  the first truly geoeffective CME of solar cycle 24 \citep{cm10,apr11}.
  Our reconstruction
  of this event has already been published \citep{bew11}, as
  well as being discussed at length in the context of our survey of
  geoeffective events \citep{bew16b}, so we do not discuss
  the reconstruction further here.  There are two overlapping
  MCs associated with this event (see Figure~1), but there is clearly
  only one CME to account for both of them.  The Earth arrival time of
  our model FR is in better agreement with the first MC, and its
  orientation is in better agreement with it as well, so we simply
  associate the FR with that MC in Table~1.  But the FR encounter time
  suggested by the rather fat reconstructed FR is more than long enough
  to encompass both MCs.
\item[Event \#9] There are two large, bright CMEs launched roughly
  towards Earth, and both are reconstructed (see Figure~6).
  The first one (CME1), launched at 16:35 on 2010~May~23, is the one we
  associate with the MC.  The faster second one (CME2) erupts
  about 21 hours later.  It decelerates to a similar final speed
  to CME1, so it never overtakes CME1.  Figure~6 shows the situation
  as the CME1 FR is hitting Earth.  Unlike CME1, CME2 has a
  very clear shock structure around it, which is modeled in our
  reconstruction as well.  Our reconstruction has the CME2 shock
  hitting Earth, but the CME2 FR misses to the west, which is why
  we associate the MC with CME1.  The CME1 arrival time shown in
  Figure~1 is about 4 hours after the ICME start time, but about 12
  hours before the MC start time.  Our reconstruction actually
  has the CME2 shock hitting Earth at about the MC start time,
  leading to the interesting question, what effect does the CME2
  shock have on CME1, and is there really any evidence for the
  CME2 shock in the in~situ data?  We do not think so, and it seems
  to us likely that the CME2 shock seen clearly in the imaging does
  not actually extend inside the FR of CME1.  This contradicts the
  reconstruction in Figure~6, but this is simply a limitation of the
  kind of reconstruction we are doing.  In multi-event reconstructions
  of this nature we can only reconstruct events independently and then
  superimpose them, which can potentially be misleading.  \citet{nl12a}
  provide a detailed assessment of the evidence
  for interaction between these two CMEs.  Our CME1 trajectory and
  FR tilt angle from Table~5 ($\lambda_s=12^{\circ}$,$\beta_s=3^{\circ}$,
  $\gamma_s=50^{\circ}$) can be compared with the ($\lambda_s=10^{\circ}$,
  $\beta_s=0^{\circ}$,$\gamma_s=65^{\circ}$) measurement of \citet{nl12a}.
  Finally, we note that with $V_{max}=385.9$ km~s$^{-1}$,
  CME1 is the slowest flare-associated CME in our sample, the only
  flare-associated event that never exceeds 500 km~s$^{-1}$.
\item[Event \#10] This CME is notable for early rotation \citep{av11},
  and for hitting the {\em MESSENGER} spacecraft
  in addition to hitting Earth \citep{tnc12}.  The
  time-dependent behavior noted by \citet{av11} means
  self-similar expansion is not a very good approximation close to
  the Sun.  Thus, our reconstruction fits the COR1 data poorly, as we
  have focused more on reproducing the COR2 and HI1 appearance well.
  There is a CME that erupts about 10 hours after our event.  We
  believe this CME misses Earth to the east, and in our judgment
  remains separated from our event to the extent that we do not
  do a reconstruction for it, but this second CME confuses
  interpretation of the SOHO/LASCO images somewhat.
\item[Event \#11] This is an extensively analyzed event, popular
  because of its geoeffectiveness, and for studying CME-CME
  interaction \citep{cjs11b,ccw11,mt12,rah12,cm12,dfw13}.
  However, from an imaging perspective, this is
  a frustrating event, because data gaps mean there is no
  available SOHO/LASCO data, and STEREO-B data gaps
  lead to minimal COR2-B and HI1-B constraints.  For our reconstruction,
  we ultimately decide to focus on only two of the 5 or 6 events
  discussed in the literature cited above:  CME2 and CME3 launched
  at about 8:00 and 8:30 on 2010~August~1, respectively.  There is
  an earlier event (CME1) from about 3:10 that is also relevant for
  this discussion, which appears to be swept up and obliterated by
  the faster, brighter CME2, early in the HI1-A field of view.
  Figure~6 shows the situation as the shock of CME2 approaches Earth.
  It is the CME2 shock that is responsible for the shock that marks
  the ICME start time in Figure~1.  Our reconstruction has the shock
  arriving about 5 hours too early.  However, this reconstruction has
  the FR of CME2 missing Earth entirely, meaning we ultimately do
  not associate CME2 with either of the 2 consecutive MCs shown in
  Figure~1.  Our reconstruction has the FR of CME3 reaching {\em Wind}
  near the start time of the second MC, so it is the second MC that
  we connect with CME3 in Table~1.  The orientation of MC\#2 is
  also in better agreement with the FR orientation than is the
  case for MC\#1.  It is possible that the short MC\#1 may be
  associated with the swept up remnants of CME1, as suggested by
  \citet{cm12}.
\item[Event \#12] The FR encounter time implied by the reconstruction
  is much longer than the MC time, but there are density peaks
  relatively close to both the FR entry and exit times.
\item[Event \#13] This event looks like a nice north-south oriented
  FR CME, somewhat analogous to event \#24 shown in Figure~3.
  The predicted ICME arrival time at {\em Wind} nicely matches an
  observed density peak at that time (see Figure~1), but we have
  assumed an ICME arrival time about 14 hours earlier, where a
  velocity jump is observed along with a smaller density increase.
  Complicating interpretation further is that the MC start time
  is actually about 12 hours {\em after} the predicted ICME arrival.
  This illustrates the kinds of difficulties that can arise when trying
  to connect a front followed in images with features in in~situ data.
\item[Event \#14] This is a CME with a particularly slow, gradual
  initial expansion.  This event and event \#2 are the only events
  in our sample that take over 50 hours to reach their peak speeds of
  $V_{max}\approx 400$ km~s$^{-1}$, by which time they are both well
  into the HI1 field of view.
\item[Event \#15] This event has garnered significant attention
  for being associated with the first X-class flare of solar
  cycle 24, and for providing another interesting example of CME-CME
  interaction \citep{cjs11a,dm14,wm14,mt14}.  Our reconstruction
  involves two CMEs: a slow CME (CME1) that begins at about 16:24 on
  2011~February~14, followed by a faster CME (CME2) at 1:55 on
  2011~February~15, which is the one associated with the X-flare.
  It is only CME2 that ultimately hits Earth in our reconstruction.
  It catches up with CME1 in the HI1 field of view, and seems to nudge
  it further to the north than it already was.  With a peak speed of
  $V_{max}=1387.8$ km~s$^{-1}$, CME2 ends up being the fastest CME in
  our sample, and has a visible shock that is modeled in the
  reconstruction.  The kinematic model for this event in Figure~4(b) is
  unusual for having two deceleration phases instead of just one.
\item[Event \#16] This is the final slow, streamer blowout type CME in
  our sample, with no clear surface activity associated with it.
  A faint but complete halo is apparent in the SOHO/LASCO images.
  Self-similar expansion is not a terribly good approximation for
  this event, limiting the degree to which our model can reproduce
  the CME appearance in the images.
\item[Event \#17] Our reconstruction involves two CMEs, the first (CME1)
  starting at 4:20 on 2011~May~25, and the second (CME2) starting at
  about 12:50.  Both are quite faint in COR1 and in LASCO.  The faster
  CME2 catches up to and slightly overtakes CME1 in the HI1 field of
  view.  Figure~6 shows the situation as CME1 is reaching Earth.  Our
  reconstruction has CME2 just barely missing Earth to the west, but
  this is a case where there is significant doubt as to which of the
  two CMEs is responsible for the MC.  The big velocity increase after
  the MC period in Figure~1 suggests the CME may have arrived at Earth
  in the midst of a CIR.
\item[Event \#18] This event was analyzed in the previous survey of
  cycle 24 geoeffective events \citep{bew16b}.  Our
  reconstruction is described in detail in this previous work.  There
  are three modeled CMEs in this difficult reconstruction (CME1, CME2,
  CME3), all emanating from the same active region, with the three
  CMEs seeming to merge in the HI1 field of view.  There is some
  uncertainty regarding whether CME2 and/or CME3 are Earth-directed,
  but our reconstruction ultimately
  suggests that only CME3, the fastest and largest of the three CMEs,
  encounters Earth, and it only just barely grazes it.  There are
  two consecutive MCs identified in the {\em Wind} data.  We
  tentatively identify the CME with the longer second one, as the
  FR orientation seems more consistent with this one.
\item[Event \#19] Our reconstruction has the Earth just barely grazing
  the bottom of the west leg of this CME.  This event and event \#24
  shown in Figure~3 have several interesting characteristics in
  common.  Both are associated with filament eruptions from very large
  filament channels, but no flares.  There is no impulsive
  acceleration, but both gradually accelerate to surprisingly high
  speeds of $V_{max}=798.9$ km~s$^{-1}$ and $V_{max}=907.6$
  km~s$^{-1}$, respectively; and they do not decelerate, making these
  the only CMEs in our sample with $V_{max}>600$ km~s$^{-1}$ that
  never decelerate.
  Finally, these are the broadest FRs
  in our sample, with $FWHM_s=119.8^{\circ}$ and
  $FWHM_s=118.4^{\circ}$, respectively.  There is some suggestion that
  a shock develops ahead of event \#19 in the HI1 field of view,
  particularly visible in HI1-A.  If true, this may be an ideal case for
  studying shock development in the interplanetary medium, which
  happens slowly and far from the Sun for this event thanks to its
  relatively slow, gradual acceleration.
\item[Events \#20-21] There is a clear SDO/AIA brightening and post
  eruption loops associated with this event in an active region (AR
  11289) near disk center, but the C2.9 flare taking place at about
  this time may be due to simultaneous activity near the east limb.
  We call this a flare-associated CME, but do not give a GOES
  designation in Table~1 due to this ambiguity.  This is not the end
  of the ambiguity for this event, unfortunately.  There are two
  consecutive MCs associated with the {\em Wind} ICME (see Figure~1),
  and close inspection of the imaging leads us to conclude that there
  are in fact two distinct CMEs erupting simultaneously from the same
  active region, which are very hard to separate.  Our reconstruction
  therefore includes two Earth-directed CMEs launched simultaneously,
  CME1 and CME2.  CME1 is somewhat faster and overtakes the CME2 front
  in COR1.  CME1 is directed $\lambda_s=23^{\circ}$ west of
  Earth, while CME2 is at $\lambda_s=-2^{\circ}$ and with a more
  north-south orientation.  Perhaps CME2 is associated with a part of
  the eruption above the post-eruption loops, while CME1 is more
  associated with an EUV dimming region west of the active region.  In
  LASCO/C2, the combined CME is seen as a partial halo, with the part
  of the halo to the west and northwest expanding faster than the part
  due north, a further indication that this is actually two distinct
  eruptions (CME1 and CME2, respectively).  From STEREO-B's
  perspective the two CMEs end up almost perfectly superimposed after
  COR1, but separation between the leading edges of the two CMEs
  is apparent for STEREO-A.  Finally, there is a jet of material
  towards the west (CME3), which has been modeled in the
  reconstruction as a lobular front (not an FR), although it comes
  nowhere near Earth.  The order of arrival at Earth is CME1 hitting
  first and CME2 following only a couple hours later.  The predicted
  encounter times of both are easily long enough to encompass both MC
  periods, so it is dubious whether we can really connect the two CMEs
  with the two MCs, but for purposes of our study an association is
  required.  Both the arrival time order and FR orientation are
  suggestive of CME1 and CME2 being associated with the first and
  second MCs, respectively, so that is the assumption made in Tables~1
  and 2.  In the {\em Wind} data, near the ICME start time, there is a
  very narrow high density peak, which also has magnetic field
  signatures.  It is tempting to associate this very unusual in~situ
  signature somehow with the complex nature of this event, with
  apparent overlapping and interacting CMEs.
\item[Event \#22] Our analysis involves the reconstruction of four
  separate CMEs: CME1 beginning at 9:20 on 2011~October~1, CME2 at
  10:00, CME3 at 20:40, and CME4 at 1:00 on October~2.  Neither CME1
  nor CME3 are Earth-directed, with the former being directed west of
  Earth and the latter directed well to the east, more towards
  STEREO-B.  CME3 is by far the fastest and brightest of the CMEs,
  with a clear shock that is included in the reconstruction.  This CME
  is already mostly out of the field of view at the time shown in
  Figure~6.  Our reconstruction has both CME2 and CME4 hitting Earth.
  The faster CME4 overtakes CME2 late in the HI1 field of view and
  ultimately hits Earth first, at a time close to the ICME and MC
  start time (see Figure~1), so this is the CME that we associate with
  the MC.  Figure~6 shows the situation at about the time that CME2 is
  hitting Earth.  There is always going to be some ambiguity in
  interpretation when two CMEs overlap like this, though.  Could CME4
  have accelerated the CME2 FR, so that it is still CME2 that accounts
  for the MC?  Did CME4 deflect CME2 slightly to the east, meaning
  CME2 misses Earth entirely, which our reconstruction cannot
  reproduce due to the simple radial, self-similar expansion
  assumption?  We prefer this latter interpretation, but we note that
  the predicted CME2 arrival time corresponds to the time of a big
  density increase near the end of the MC period in Figure~1, so
  perhaps that is the CME2 signature.
\item[Event \#23] This reconstruction involving three CMEs is described
  in detail by \citet{bew16b}, so we do not discuss it in great
  detail here.  Only the second CME to erupt hits Earth.  This is
  the most geoeffective CME in our sample, as measured by $D_{st}$
  (see Table~2), despite being of only modest speed
  ($V_{max}=720.3$ km~s$^{-1}$), and not being associated with a flare.
\item[Event \#24] As the only $Q_s=5$ event in our sample, we have
  already discussed this event in Section~3.2 (see Figure~3), and also
  above in the discussion of event \#19, which resembles this event in
  many ways.  With regards to its in~situ signature, this is an
  instructive event for illustrating the ambiguities there can be
  connecting imaging and in~situ signatures of a CME.  We see nothing
  in the images that we associate with a shock, so the leading edge
  that we follow in the images is interpreted as the leading edge of
  the FR.  The predicted Earth-arrival time agrees very well with the
  ICME start time marked in Figure~1, but the {\em Wind} in~situ data
  {\em does} show a probable shock at that time, with the MC actually
  starting about 13 hours later.  Does this mean that what we were
  following in the images was a shock all along?  Did that front
  suddenly become a shock at some point in its interplanetary
  journey without us noticing?  Given that our reconstruction has
  Earth encountering the southern leg of the FR, does this mean that
  there is more of a shock on the lateral parts of the FR than near
  the leading edge?  There is not a very big density increase at the
  start of the MC that would suggest that we would have been following
  that in the images.  It is notable that the predicted velocity is
  much higher than the observed velocity for the period between ICME
  arrival and the MC start time.  This is to be expected for a shock
  and shock sheath, as images will track the shock propagation speed,
  which will be faster than the speed of material moving through the
  shock, which is what the in~situ data will be measuring.  It is also
  instructive to note that within the MC region the predicted and
  observed velocity curves agree well, consistent with this being CME
  ejecta (i.e., the FR).
\item[Event \#25] Our reconstruction has Earth encountering the north
  leg of a north-south oriented FR.  This CME is only just barely
  trackable into the HI2 field of view.  The {\em Wind} data do not
  show very significant density, velocity, or field increases for
  this MC.
\item[Event \#26] Our reconstruction includes a shock, which is
  clearly visible ahead of this bright, fast CME
  ($V_{max}=1103.5$ km~s$^{-1}$).  This is easily the fattest FR in
  our entire sample, with $\Lambda_s=0.43$.  This is a case where the
  lack of a clear trailing edge to the FR prevents this from being a
  very convincing case for an FR morphology, hence $Q_s=2$ in Table~2.
  A spheroidal blob model might have worked just as well.
  Our Earth-arrival time prediction is nicely consistent with the MC
  start time, but the {\em Wind} data show a much earlier arrival for
  the shock than our reconstruction predicts.
\item[Events \#27-28] There are many CMEs occurring in this time
  period, greatly complicating interpretation of the images.  For our
  reconstruction, we decide in the interests of time to focus only on
  the two CMEs that we ultimately decide are Earth-directed: CME1
  starting at 20:45 on 2012~July~3 and CME2 starting at 16:45 on
  July~4.  These events are not very apparent from the frontal
  perspective of SOHO/LASCO.  The brightest, fastest fronts seen
  passing over Earth in HI2-A on July~6 and 7 are other CMEs, not CME1
  and CME2, only hints of which can be seen moving more slowly in the
  background.  The two CMEs of interest are invisible in HI2-B.  The
  difficulty seeing CME1 and CME2 in HI2 is indicative of the
  worsening viewing geometry for Earth-directed CMEs in mid-2012,
  which is why we did not continue to analyze events after 2012~July.
  Figure~6 shows the CME positions as CME1 is reaching Earth.  We
  tentatively associate CME1 and CME2 with the first and second MCs
  identified in the {\em Wind} data, respectively (see Figure~1).  The
  CME1 shock arrives at Earth in good agreement with the ICME start
  time, and the CME1 FR arrival agrees well with the start of the
  first MC.  The arrival time of CME2 corresponds reasonably well with
  the beginning of the second MC.
\end{description}

\end{document}